\documentclass{article}

 \usepackage[preprint]{neurips_2026}

\usepackage[utf8]{inputenc} % allow utf-8 input
\usepackage[T1]{fontenc}    % use 8-bit T1 fonts
\usepackage{hyperref}       % hyperlinks
\usepackage{url}            % simple URL typesetting
\usepackage{booktabs}       % professional-quality tables
\usepackage{amsfonts}       % blackboard math symbols
\usepackage{nicefrac}       % compact symbols for 1/2, etc.
\usepackage{microtype}      % microtypography
\usepackage{xcolor}         % colors

\usepackage{amsmath}
\usepackage{enumitem}
\usepackage{graphicx}
\usepackage{wrapfig}
\usepackage{subcaption}
\usepackage{wrapfig}
\usepackage{hyperref}
\title{Prototypes of the Mind: A Unified Framework for Probing the Visual Brain}
\author{
    Shi Chen\textsuperscript{1} \qquad
    Seoyoung Ahn\textsuperscript{2} \qquad
    Doris Tsao\textsuperscript{1} \\[0.5em]
    \textsuperscript{1}Department of Neuroscience,
    UC Berkeley \\
    \textsuperscript{2}College of AI Convergence, 
    Hankuk University of Foreign Studies
}

\begin{document}

\maketitle

\begin{abstract}
Deep learning has driven substantial progress in modeling fMRI responses, yet most existing approaches are optimized for a single objective---such as neural encoding, stimulus decoding, or post hoc interpretation---making it difficult to obtain a coherent account of neural representations. We introduce a unified framework that connects vision foundation models to brain activity while supporting neural analysis from multiple complementary perspectives. Our approach replaces black-box features with a sparse, interpretable representation composed of visual prototypes, and links these prototypes to voxel responses through a linear readout. This design makes the model transparent at both levels: prototype activations expose the visual concepts present in the stimulus, while the learned linear weights characterize each voxel's selectivity. Using this interface, we identify which visual concepts drive neural responses and where they appear in the image, reveal cortical organization through voxel tuning in prototype space, and perform controlled stimulus synthesis to probe cortical selectivity through targeted interventions. Together, these results establish prototype-based representations as a unified and interpretable foundation for studying visual representations in the human brain. Our source code is \href{https://github.com/szzexpoi/ProtoMRI}{publicly available}.

\end{abstract}

\section{Introduction}
Understanding how the human brain represents complex visual scenes remains a fundamental challenge at the intersection of neuroscience and computer vision. Neural measurements are sparse and noisy, while natural scenes are incredibly rich and complex. Modeling the relationship between the two therefore requires representations that are both predictive of neural responses and expressive enough to capture the structure of the visual world.

Advances in deep learning, together with the increasing availability of large-scale neural datasets, have transformed Functional Magnetic Resonance Imaging (fMRI) analysis by enabling increasingly effective computational models of brain responses. One major line of work develops encoding models \cite{Konkle2022_selfsupervised, Wang2023_multimodal_encoding, Yang_2024_CVPR, adeli2025transformer, Conwell2024} that seek to predict fMRI responses from visual stimuli with high accuracy. A complementary line of research tackles the inverse problem through decoding models \cite{luo2023brain, Takagi_2023_CVPR, feng2026towards, nan2026metalearningincontextenablestrainingfree, NEURIPS2023_mindeye, brainit}, which reconstruct or synthesize stimuli directly from neural activity. Moving beyond black-box features learned by deep networks, several recent studies develop interpretable representations that reveal categorical \cite{zeng2025disentangling, brainexplore} and spatial \cite{NEURIPS2023_90e06fe4, luo2025brainmappingdensefeatures} selectivity. Collectively, these approaches have produced valuable tools for probing the functional organization of the brain. However, they have largely developed in isolation, each optimized for a single question---prediction, reconstruction, or interpretation---with no shared framework. This fragmentation makes it difficult to develop a comprehensive and coherent picture of neural representations across these complementary dimensions.

In this paper, we investigate a different possibility: can we derive a single unified representation that supports the analysis of neural activity across multiple dimensions? Such a representation can serve as an interpretable bridge between vision foundation models \cite{dino, ijepa, He_2022_CVPR} and the brain, grounding neural analysis in the semantic structure learned from large-scale visual data. At the same time, by linking brain responses to a representational space with strong generative properties, it can support more advanced applications that go beyond passive analysis, including stimulus reconstruction and controlled manipulation of visual content.

Realizing this goal requires a representation that is semantically interpretable, structurally organized, and flexible enough to support both analysis and intervention. Guided by the intuition that complex visual scenes can be decomposed into distinct visual concepts \cite{chen2019protopnet, koh2020concept}, we factorize rich features from vision foundation models (\textit{e.g.,} DINOv3 \cite{dinov3}) into spatially grounded prototype units that capture recurring visual patterns in natural images. The resulting activations are sparse and preserve spatial layout, allowing each brain voxel to be associated with a small set of visual concepts using a linear projection. This yields a unified framework for brain analysis that reveals both \emph{what} drives neural activity and \emph{where} those driving concepts appear. Crucially, our compositional representation further enables \emph{targeted stimulus interventions} on interpretable prototypes (\textit{e.g.,} manipulating face and body prototypes), opening a new route to causally probe the relationship between neural activity and visual concepts.

\section{Related Work}
Recent work on fMRI modeling largely focuses on three complementary directions:

\paragraph{Encoding models.} A substantial body of work \cite{kay2008identifying, NASELARIS_survery, Wang2023_multimodal_encoding, Yang_2024_CVPR, adeli2025transformer, Konkle2022_selfsupervised, Conwell2024} learns the mappings from visual features to neural responses with encoding models. Early studies~\cite{kay2008identifying} demonstrated that voxel activity in early visual cortex can be predicted using Gabor-based features. Subsequent research expanded this framework using richer semantic or deep visual features \cite{NASELARIS_survery}, enabling more accurate prediction in higher visual areas. Recent studies increasingly leverage features from deep networks or multimodal foundation models \cite{Konkle2022_selfsupervised, Wang2023_multimodal_encoding, Yang_2024_CVPR, adeli2025transformer, Conwell2024}, which capture hierarchical visual representations aligned with cortical organization. While these DNN-based approaches are effective for predicting neural activity and characterizing voxel selectivity, they typically operate as black-box predictors, offering limited interpretability about which specific image properties drive the responses.

\paragraph{Decoding models.} Another line of work attempts to reconstruct or synthesize visual stimuli from neural activity using decoding models \cite{luo2023brain, GU2022118812, PONCE2019999, RatanMurty2021, Takagi_2023_CVPR, feng2026towards, nan2026metalearningincontextenablestrainingfree, NEURIPS2023_mindeye, brainit}. Early studies used optimization-based approaches to generate stimuli that maximally activate neurons or brain regions \cite{PONCE2019999, RatanMurty2021}. More recent methods combine fMRI decoding with powerful generative models such as diffusion models \cite{Takagi_2023_CVPR, NEURIPS2023_mindeye, brainit}, enabling reconstruction of naturalistic images from measured neural activity. These approaches provide compelling visualizations of encoded information, but they typically rely on large pretrained generative models and operate through latent spaces that are difficult to interpret or manipulate. As a result, they fall short of probing the causal relationship between neural activity and specific image properties. In contrast, our prototype representation provides a compact and interpretable feature space that supports \textit{direct manipulation of visual concepts}, enabling controlled stimulus interventions such as feature addition, removal, and spatial manipulation.

\paragraph{Spatial selectivity analysis.} Several studies investigate the spatial structure of neural responses to natural images \cite{NEURIPS2023_90e06fe4, luo2025brainmappingdensefeatures}. These approaches analyze spatial feature maps extracted from deep networks and relate them to voxel activity patterns to characterize where visual information is processed within an image. For example, recent work has shown that dense feature maps from vision models can help localize image regions driving neural responses across different cortical areas \cite{NEURIPS2023_90e06fe4}. Such analyses offer valuable insight into spatial attention and receptive field structure, but typically require separate post-hoc analysis pipelines built on top of encoding models. In contrast, our prototype representation preserves spatial structure by design, enabling spatial selectivity to emerge naturally from the model. This allows us to directly visualize how specific visual concepts contribute to neural responses.

\section{Methodology}
\begin{figure}[t]
    \centering
    \includegraphics[width=0.8\linewidth]{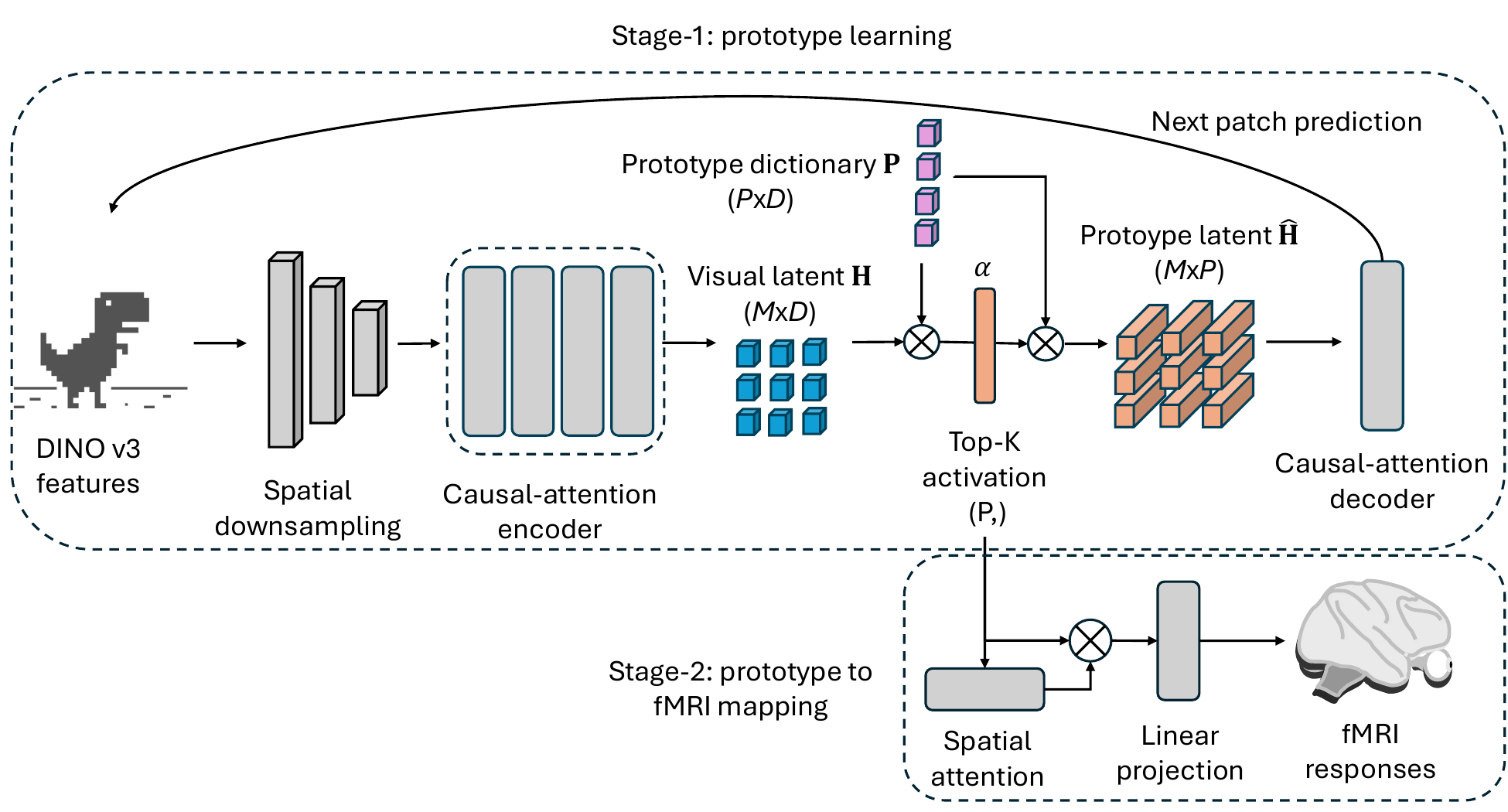}
    \caption{
        Overview of method: (1) In stage-1, we learn interpretable prototypes with a next patch prediction objective. Given patch-level features (\textit{e.g.,} from DINOv3), we factorize them as a sparse combination of learnable prototypes, which captures distinct semantics and the spatial structure of the visual scene. (2) In stage-2, we train a linear mapping between the activation of prototypes and fMRI responses, enabling brain understanding from complementary perspectives.         
    }
    \label{fig:pipeline}
\end{figure}

At the core of our method is a two-stage framework that connects visual perception with neural activity through interpretable prototypes. In Stage~1 (\S\ref{sec:stage1}), we learn a dictionary of visual prototypes that decomposes foundation model features into sparse, interpretable components. Using a self-supervised learning paradigm that emphasizes contextual relationships among visual regions, the model supports a rich interpretation of both the concepts present in a scene and their spatial organization. In Stage~2 (\S\ref{sec:stage2}), we learn a lightweight mapping from spatially distributed prototype activations to fMRI voxel responses, producing an interpretable weight matrix that directly characterizes the associations between visual prototypes and brain areas. An overview of our pipeline is shown in Figure~\ref{fig:pipeline}.

\subsection{Stage 1: Reconstructing Visual Scenes with Interpretable Prototypes}
\label{sec:stage1}

Vision foundation models such as DINO~\cite{dino} have emerged as the \emph{de facto} backbone for predicting fMRI responses~\cite{Conwell2024, adeli2025transformer, gifford2023algonautsproject2023challenge}, delivering state-of-the-art encoding accuracy across visual cortex. Yet, despite their strong predictive performance, the entangled structure of these features makes it challenging to identify which image properties drive the predicted responses in different brain regions. Our objective is to \emph{re-express} these features as sparse combinations of learnable prototypes, each grounded in concrete visual patterns, while retaining the discriminative power of the original features. To achieve this, we introduce an encoder-decoder network with three main components:

\paragraph{Causal encoder with spatial downsampling.} 
Given the patch-level features $\mathbf{X} \in \mathbb{R}^{N \times D}$ extracted from a foundation model (\textit{e.g.,} DINOv3, $N=16 \times 16$ is the number of spatial patches and $D=768$ is the feature dimension), we first apply a learnable layer \cite{yu2024language} to reduce the feature map resolution from $N = 16 \times 16$ to $M = 8 \times 8$. This step both lowers computational cost and facilitates model optimization:
\begin{equation}
    \mathbf{Z} = \texttt{SpatialDown}(\mathbf{X}) \in \mathbb{R}^{M \times D}.
\end{equation}
The downsampled features are combined with sinusoidal positional embeddings $\mathbf{E}_{pos} \in \mathbb{R}^{M \times D}$, and passed through a stack of four causal self-attention blocks $f_{enc}$ \cite{transformer, radford2018improving} for further refinement:
\begin{equation}
    \mathbf{H} = f_{\text{enc}}\!\left(\mathbf{Z} + \mathbf{E}_{\text{pos}}\right) \in \mathbb{R}^{M \times D},
\end{equation}
By preserving spatial structure in the foundation model representation, the encoder enables us to identify how image properties at different locations relate to brain responses.

\paragraph{Prototype dictionary and sparse decomposition.} 
A principal idea behind our approach is to decompose foundation model features into interpretable units. For this, we maintain a dictionary of prototypes $\mathbf{P} \in \mathbb{R}^{P \times D}$, where $P=1000$ denotes the number of prototypes and each row $\mathbf{p}_p \in \mathbb{R}^{D}$ corresponds to a single prototype. We then factorize the encoded features $\mathbf{H}$ as weighted combinations of these prototypes. For each encoded patch $\mathbf{h}_m \in \mathbb{R}^{D}$, the prototype weights $\mathbf{s}_m$ are computed from its cosine similarities with all prototypes:
\begin{equation}
    \mathbf{s}_m = \mathbf{h}_m \mathbf{P}^{\top} \in \mathbb{R}^{P}.
\end{equation}
With the overarching goal of developing an interpretable representation, we draw inspirations from sparse autoencoder for analyzing language models \cite{huben2024sparse} and further impose sparsity constraints on prototype selection. Specifically, we retain only the top-$k$ ($k=128$) prototype weights for each patch and normalize them with a softmax:
\begin{equation}
    \tilde{s}_{m,p} =
    \begin{cases}
        s_{m,p} & \text{if } s_{m,p} \in \text{top-}k(\mathbf{s}_m) \\
        -\infty & \text{otherwise}
    \end{cases}
    \qquad
    \alpha_{m,p} = \frac{\exp(\tilde{s}_{m,p})}{\sum_{j=1}^{P} \exp(\tilde{s}_{m,j})}
    \label{eq:topk}
\end{equation}
The resulting prototype activation vector $\boldsymbol{\alpha}_m \in \mathbb{R}^{P}$ preserves exactly the $k$ most relevant prototypes as nonzero entries while suppressing the remaining $P-k$ prototypes. Each patch is then reconstructed as a weighted combination of prototype vectors:
\begin{equation}
    \hat{\mathbf{h}}_m = \sum_{p=1}^{P} \alpha_{m,p} \cdot \mathbf{p}_p \in \mathbb{R}^{D}.
    \label{eq:proto_recon}
\end{equation}

Stacking over all $M$ patches, we obtain the prototype-decomposed representation $\hat{\mathbf{H}} = \mathbf{A}\mathbf{P} \in \mathbb{R}^{M \times D}$, where $\mathbf{A} = [\boldsymbol{\alpha}_1, \ldots, \boldsymbol{\alpha}_M]^{\top} \in \mathbb{R}^{M \times P}$. Each row of $\mathbf{A}$ indicates, for a given spatial token, the strength with which each of the $P$ prototypes is activated, while each column shows how a particular prototype is used across the $M$ spatial tokens. Together, $\mathbf{A}$ provides a sparse, interpretable decomposition of the visual scene in terms of prototypes and their spatial distribution.

\paragraph{Causal decoder and next-patch prediction.}
The reconstructed representation $\hat{\mathbf{H}}$ is passed through a decoder composed of a single causal self-attention block followed by a linear readout head:
\begin{equation}
    \hat{\mathbf{Y}} = g_{\text{read}}\!\left(f_{\text{dec}}(\hat{\mathbf{H}})\right) \in \mathbb{R}^{M \times D'},
\end{equation}
Note that the readout projects each decoded patch into a higher-dimensional space ($D'{=}4D{=}3072$), to accommodate the $2{\times}2$ spatial downsampling within the encoder. 

To preserve the discriminative power of the learned prototypes, we adopt an autoregressive training paradigm used in large language \cite{radford2018improving} and multimodal \cite{gupta2026better} models. Specifically, we rasterize the visual patches, and task our model to predict features of the next patches based on preceding ones:
\begin{equation}
    \mathcal{L}_{proto} = \frac{1}{M-1} \sum_{m=1}^{M-1} \left\| \hat{\mathbf{y}}_m - \mathbf{y}_{m+1} \right\|^2,
    \label{eq:loss_proto}
\end{equation}
where $\hat{\mathbf{y}}_m$ is the prediction at position $m$ and $\mathbf{y}_{m+1}$ is the target at position $m{+}1$. This objective encourages the model to capture the spatial structure without the need of additional annotations.

\subsection{Stage 2: Bridging Visual Scenes and the Brain with Prototypes}
\label{sec:stage2}
Stage~1 produces, for each image, a sparse activation map $\mathbf{A} \in \mathbb{R}^{M \times P}$ where $M{=}64$ spatial locations each activate at most $k=128$ prototypes. We now leverage this representation to model fMRI responses. Instead of developing a black-box encoding model \cite{Yang_2024_CVPR, adeli2025transformer, Konkle2022_selfsupervised, Conwell2024} that poses challenges for interpretation, our mapping from prototypes to neural activity is kept simple: a learned spatial attention aggregates prototype activations across locations, and a single linear layer projects the aggregated prototype vector onto voxel responses. The combination of interpretable prototype features with a linear encoding model yields a transparent interface between visual content and neural activity---each weight directly quantifies how strongly a given prototype drives a given voxel.

\paragraph{Spatial aggregation via attention.}
As different brain areas may preferentially attend to different spatial locations (\textit{e.g.,} face-selective areas may emphasize central patches), we learn a spatial attention mechanism that aggregates the $M{=}64$ patch-level activations into a single image-level prototype vector. Given the sparse prototype activation map $\mathbf{A} \in \mathbb{R}^{M \times P}$, we reshape it into a 2D feature map $\mathbf{A}_{2D} \in \mathbb{R}^{P \times 8 \times 8}$, and compute the attention map with a convolutional layer:
\begin{equation}
    \mathbf{w} = \texttt{softmax}\!\left(\texttt{Conv2d}_{3 \times 3}(\mathbf{A}_{2D})\right) \in \mathbb{R}^{M \times 1},
    \label{eq:spatial_att}
\end{equation}
The spatially aggregated prototype vector per image is then:
\begin{equation}
    \bar{\boldsymbol{\alpha}} = \sum_{m=1}^{M} w_m \, \boldsymbol{\alpha}_m \in \mathbb{R}^{P},
    \label{eq:agg}
\end{equation}
where $w_m$ is the learned attention weight for spatial position $m$ and $\boldsymbol{\alpha}_m$ is the $m$-th row of $\mathbf{A}$.

\paragraph{Prototype-to-voxel linear mapping.}
The aggregated vector $\bar{\boldsymbol{\alpha}}$ is mapped to voxel responses via a linear transformation:
\begin{equation}
    \hat{\mathbf{r}} = \mathbf{W} \bar{\boldsymbol{\alpha}} + \mathbf{b} \in \mathbb{R}^{V},
    \label{eq:linear_map}
\end{equation}
where $\mathbf{W} \in \mathbb{R}^{V \times P}$ is the learned weight matrix and $\mathbf{b} \in \mathbb{R}^{V}$ is the bias.
Each row $\mathbf{w}_v \in \mathbb{R}^{P}$ of $\mathbf{W}$ encodes the sensitivity of voxel $v$ to each of the $P$ prototypes, providing a direct and interpretable characterization of voxel selectivity. The model is trained with a composite loss combining mean squared error $\mathcal{L}_{MSE}$ and Pearson's correlation $\mathcal{L}_{corr} = 1-\rho$:
\begin{equation}
    \mathcal{L}_{fMRI} = \mathcal{L}_{MSE} + \mathcal{L}_{corr}
    \label{eq:loss_fmri}
\end{equation}
both of which are commonly used for fMRI encoding models \cite{gifford2023algonautsproject2023challenge, adeli2025transformer}.

%%%%%%%%%%%%%%%%%%%%%%%%%%%%%%%%%%%%%%%%%%%%%%%%%%%%%%%%%%%%%%%%%%%%%%%%%%%%%%%%
\section{Experiments} \label{result_section}
\subsection{Datasets}
We use MSCOCO dataset \cite{lin2014microsoft} for learning our prototypes, which includes images with rich semantics. For the fMRI data, we adopt the processed NSD dataset \cite{nsd_dataset} from Algonauts 2023 \cite{gifford2023algonautsproject2023challenge}. The dataset provides annotations on the functional selectivity of different brain areas (\textit{e.g.,} Occipital Face Area selected for faces, and Extrastriate Body Area selected for bodies), and we choose 12 brain areas with distinct selectivity (\textit{i.e.}, faces, bodies, places, and words) as our testbed to verify the effectiveness of our approach. See Appendix \ref{sup:model_details} for details.

\subsection{Neural Encoding Performance}
\label{res:performance}

\begin{table}[b]
\centering
\resizebox{\linewidth}{!}{
\begin{tabular}{lcccccccc}
\hline
 & OFA & FFA-1 & EBA & FBA-1 & OPA & PPA & OWFA & VWFA-1 \\
\hline
Noise-ceiling & 0.561  & 0.597 & 0.600 & 0.541 & 0.588 & 0.563 & 0.530 & 0.551\\
Adeli \textit{et al.}\cite{adeli2025transformer} & 0.430 & 0.489 & 0.481 & 0.432 & 0.438 & 0.485 & 0.399 & 0.430\\
\hline
JEPA-VIT & 0.305 & 0.381 & 0.381 & 0.348 & 0.348 & 0.414 & 0.244 & 0.317\\
Proto JEPA-VIT & 0.355 & 0.403 & 0.404 & 0.359 & 0.399 & 0.419 & 0.314 & 0.350\\
\hline
DINOv3-ConvNeXt & 0.273 & 0.366 & 0.379 & 0.319 & 0.324 & 0.377 & 0.209 & 0.304\\
Proto DINOv3-ConvNeXt & 0.335 & 0.410 & 0.42 & 0.347 & 0.378 & 0.408 & 0.268 & 0.342\\
\hline
DINOv3-ViT & 0.302 & 0.388 & 0.396 & 0.330 & 0.332 & 0.386 & 0.254 & 0.314 \\
Proto DINOv3-ViT & 0.367 & 0.434  & 0.433 & 0.387 & 0.404 & 0.443  & 0.310 & 0.370 \\
\hline
\end{tabular}
}
\vspace{0.5em}
\caption{Comparison of fMRI encoding across 8 subjects. We report the Pearson's correlation for 8 distinct brain ROIs, averaged across all voxels. See Appendix \ref{sup_encoding_sec} for results of other ROIs.}
\label{tab:roi_correlation}
\end{table}

We evaluate how well our interpretable prototype representations predict fMRI responses across three visual backbones: JEPA-ViT \cite{ijepa}, DINOv3-ViT, and DINOv3-ConvNeXt \cite{dinov3}. These backbones allow us to assess the effectiveness of our method across different training approaches and architectures. To contextualize our results in terms of achievable encoding performance, we also report the noise ceiling and the performance of \cite{adeli2025transformer}, which uses a transformer prediction network on DINOv2 backbone features. All remaining models use linear prediction, applied either directly to the visual backbone features or to our interpretable prototype representations derived from those features.

Our prototype representations consistently improve prediction across all three backbones and all eight ROIs (Table~\ref{tab:roi_correlation}). With DINOv3-ViT, the mean correlation across ROIs increases from 0.338 to 0.394, reaching $70\%$ of the noise ceiling and $88\%$ of the correlation achieved by the nonlinear encoding model. These results show that our prototype representation captures substantial neural response information through a simple linear mapping and improve prediction over the original backbone features across different models. We further examine the effects of prototype number and top-$k$ sparsity in Appendix~\ref{sup:hyperparameter}.

The prototype representation also makes the encoding model directly interpretable: each voxel’s predicted response is a weighted combination of prototype activations, allowing us to examine which visual concepts contribute to its response. This structure supports analyses of categorical selectivity, cortical organization, and spatial selectivity, as well as controlled stimulus manipulations. In the following sections, we use the DINOv3-ViT prototype model to investigate these properties.

\subsection{Prototypes Capture Semantically Coherent Visual Concepts}
\label{sec:prototype-semantic}

\begin{figure*}[!]
    \centering
    \includegraphics[width=0.95\linewidth]{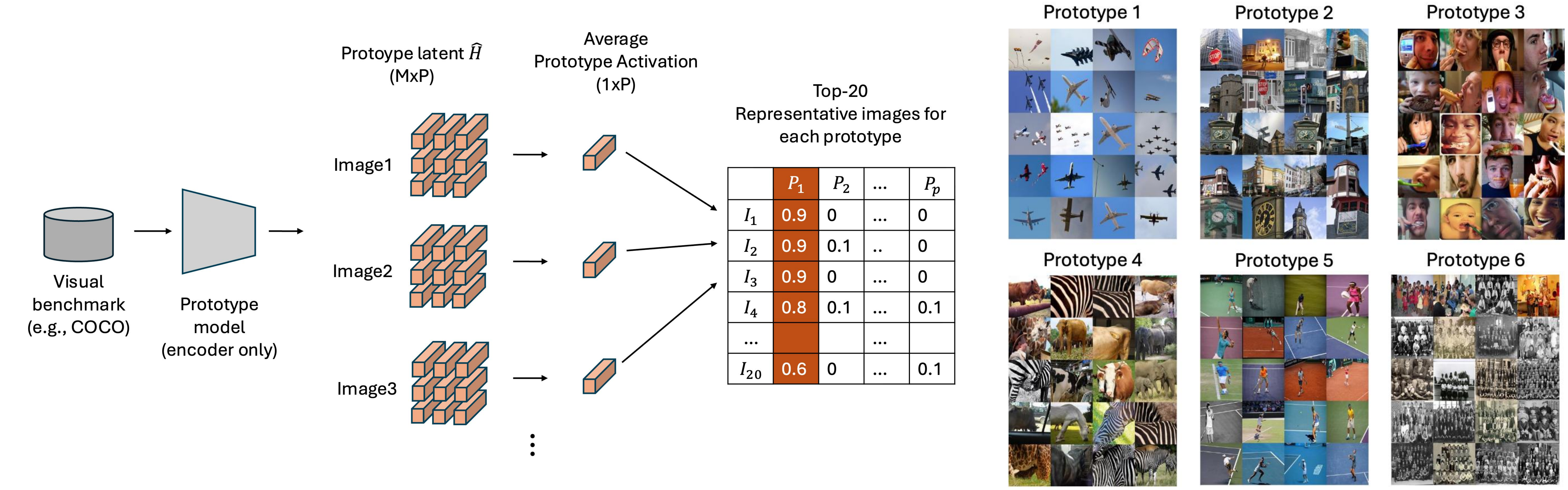}
    \caption{
    Visualization of learned prototypes. For each prototype, images are ranked by their average prototype activation (Left).
    We show representative examples among the top-activated images for several prototypes (Right). For visualization, we select 20 representative examples from each set. Additional visualization is provided in Appendix \ref{sup:visualize-more-prototypes}.
    }
    \label{fig:visualize-prototype}
\end{figure*}

We first examine whether the learned prototypes correspond to interpretable visual concepts rather than arbitrary directions in feature space. Specifically, we extract prototype activation maps $\mathbf{A}$ of images from MSCOCO validation set, and identify the representative images with strong activation (averaged across all patches) for each prototype (see Appendix \ref{sup:model_details} for details). To enhance consistency among the final representatives, we also include an optional step that computes pairwise cosine similarities between CLIP~\cite{clip} embeddings of these candidates and applies spectral clustering~\cite{ng2001spectral} to select 20 images that are both highly activated and representative of a shared visual concept (see Appendix \ref{sup:clustering_effect} for comparison between visualization with and without clustering).

As shown in Figure~\ref{fig:visualize-prototype}, individual prototypes consistently capture coherent visual concepts. Some prototypes specialize in inanimate objects (``airplanes'', ``buildings''), while others correspond to animate entities (``faces'', ``zebra'') or activities (``tennis'') or abstract scene-level concepts (``crowds''). This structure emerges entirely from the self-supervision, without explicit semantic labels, yet the learned prototypes organize images according to meaningful visual themes. On the other hand, applying the same pipeline to PCA features fails to capture meaningful semantics, indicating that our approach is not tuned for trivial image statistics (see Appendix \ref{sup:pca-roi-profile}).

\subsection{Prototypes Reveal Categorical Selectivity of Visual Cortex}
\label{res:brain-roi-profile}

Having established that the prototypes correspond to interpretable visual concepts, we next study whether their mapping to neural activity reflects the known categorical selectivity of the visual cortex. 

As our model employs a linear mapping between prototypes and fMRI responses, the weight matrix $\mathbf{W} \in \mathbb{R}^{V \times P}$ can provide a direct readout of their relationship: each voxel is associated with a set of weights indicating which prototypes most strongly drive its activity. By aggregating these weights across voxels within predefined regions of interest (ROIs), we obtain a prototype profile that summarizes the relative contribution of each prototype to a given brain area.

\begin{figure}[h]
    \centering
    \includegraphics[width=0.95\linewidth]{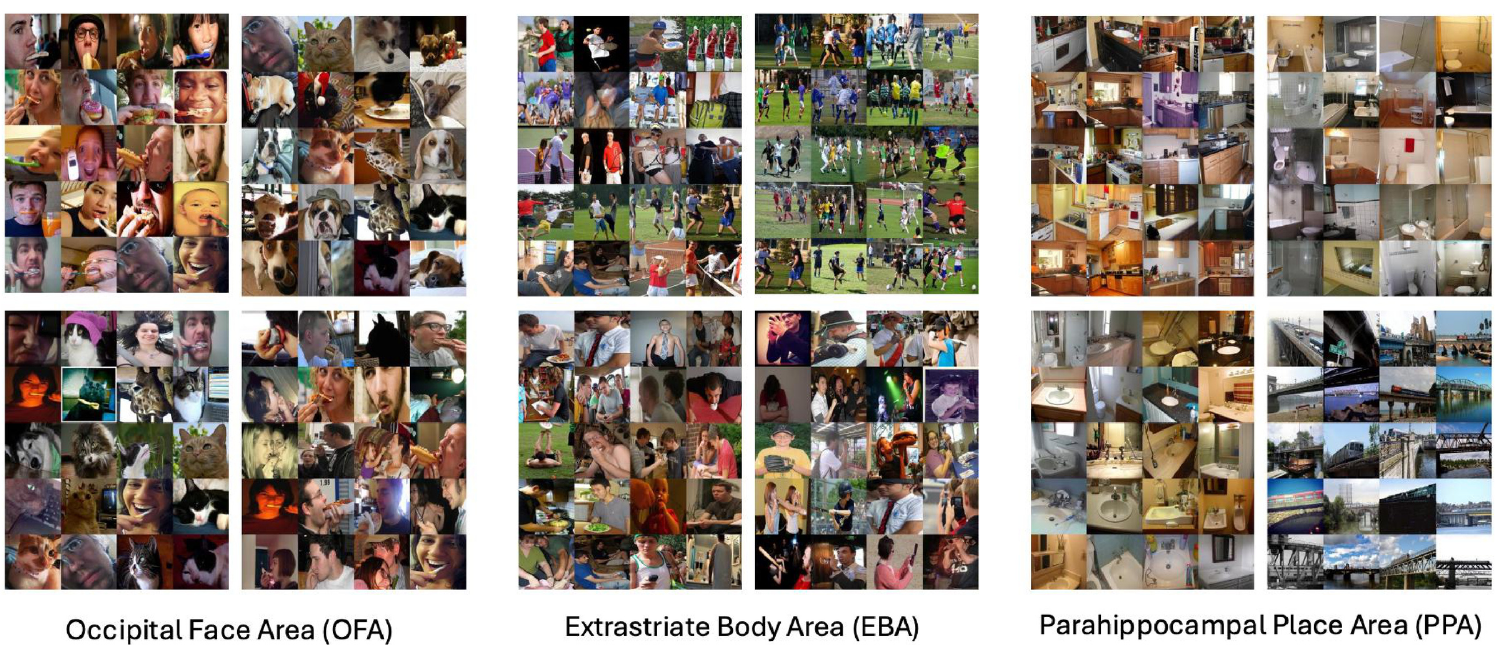}
    \caption{
    Prototype profiles (visualized as top-4 prototypes with the highest weights) for category-selective regions of the visual cortex. The profiles closely align with the expected selectivity of diverse brain areas. See Appendix \ref{sup:more-roi-profile} and \ref{sup:roi-profile-spectrum} for additional results.}
    \label{fig:roi-profile}
\end{figure}

We analyze these profiles for several well-established category-selective regions in the ventral visual pathway, including face-selective areas (e.g., Occipital Face Area, OFA), body-selective areas (Extrastriate Body Area, EBA), and scene-selective areas (Parahippocampal Place Area, PPA). If the learned representation captures meaningful visual structure, prototypes corresponding to faces, bodies, or scenes should preferentially contribute to the corresponding ROIs. Figure~\ref{fig:roi-profile} shows that the resulting prototype profiles closely align with these expected pattern, and these top prototypes also capture a considerable proportion of variance for the ROIs (see Appendix \ref{sup:prototype_predictivity}). These patterns emerge without any explicit supervision for functional localization, suggesting that the learned prototype representation captures visual features that correspond naturally to the selectivity of high-level visual cortex. As a control, we repeat the same ROI–feature analysis using PCA. Results show that it fails to characterize the selectivity of different ROIs, indicating that the ROI selectivity observed above is not a consequence of dimensionality reduction but arises from the learned prototype structure. We include the control results in Appendix \ref{sup:pca-roi-profile}. In addition, our approach also outperforms a standard sparse autoencoder in localizing semantic structures and characterizing selectivity (see Appendix \ref{sup:sae_proto_comparison}).

We also examine whether independently fitted encoding models select similar prototypes across subjects (see Appendix \ref{sup:shared-prototypes}). The substantial overlap in prototypes for the same brain areas, especially face- and body-selective regions, suggests that our prototypes generalize well across subjects.

\subsection{Mapping Prototypes to Brain Areas}
\label{res:localization}

\begin{figure}[h]
    \centering
    \includegraphics[width=0.95\linewidth]{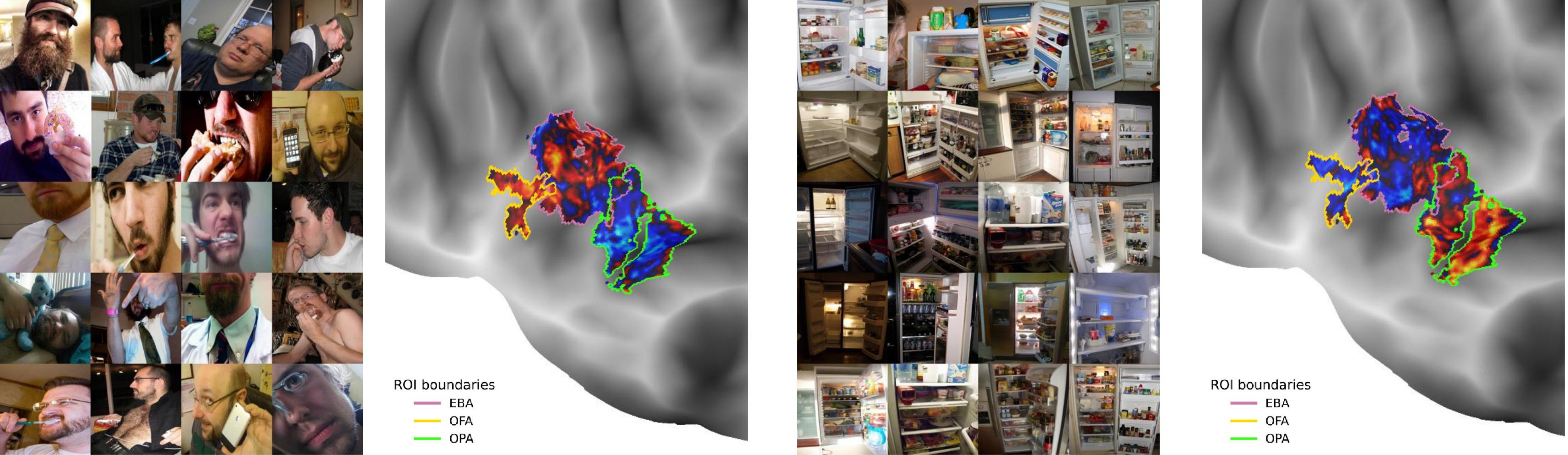}
    \caption{
    Per-prototype cortical activation maps. Strong and weak activations are denoted with yellow and blue colors, respectively. See Appendix \ref{sup:more-proto-flatmap} for additional results.
    }
    \label{fig:proto-flatmap}
\end{figure}

While the previous analysis examines brain regions and asks which prototypes they prefer, we can also ask the reverse question: for a given prototype, where in the brain is it most strongly represented? This analysis is enabled by the linear projection matrix $\mathbf{W} \in \mathbb{R}^{V \times P}$ that maps prototype activations to voxel responses. Each column $\mathbf{w}_p \in \mathbb{R}^{V}$ therefore describes how strongly prototype $p$ contributes to the predicted response of each voxel. Figure~\ref{fig:proto-flatmap} visualizes these weights on a cortical flatmap. Positive weights (yellow) indicate voxels whose activity is positively driven by the prototype, whereas near-zero or negative weights (blue) indicate weak or suppressed contributions. This analysis is conceptually related to \cite{Huth2012ContinuousSemanticSpace}, which maps semantic category features onto cortical responses using linear encoding models. Here, the feature space is instead learned directly from self-supervision.

The resulting maps reveal spatially localized and functionally meaningful activation patterns. A face-selective prototype, whose top-activated images depict human faces, produces elevated weights concentrated in face-selective cortex, particularly the occipital face area (OFA) and nearby ventral regions, while place-selective areas such as the occipital place area (OPA) show minimal response. Conversely, a kitchen-selective prototype whose representative images depict kitchen interiors produces strong activation in OPA with little contribution to OFA or EBA.

These results highlight another advantage of our framework: the combination of a simple linear decoder and semantically meaningful prototype units enables direct visualization of how specific concepts map onto cortical responses. This may help generate hypotheses about stimulus--response relationships in specific brain regions and guide the design of targeted neuroscience experiments.

\subsection{Prototypes Reveal Spatial Selectivity of Brain Areas}
\label{res:spatial-attention}

\begin{figure}
\centering
\includegraphics[width=0.95\linewidth]{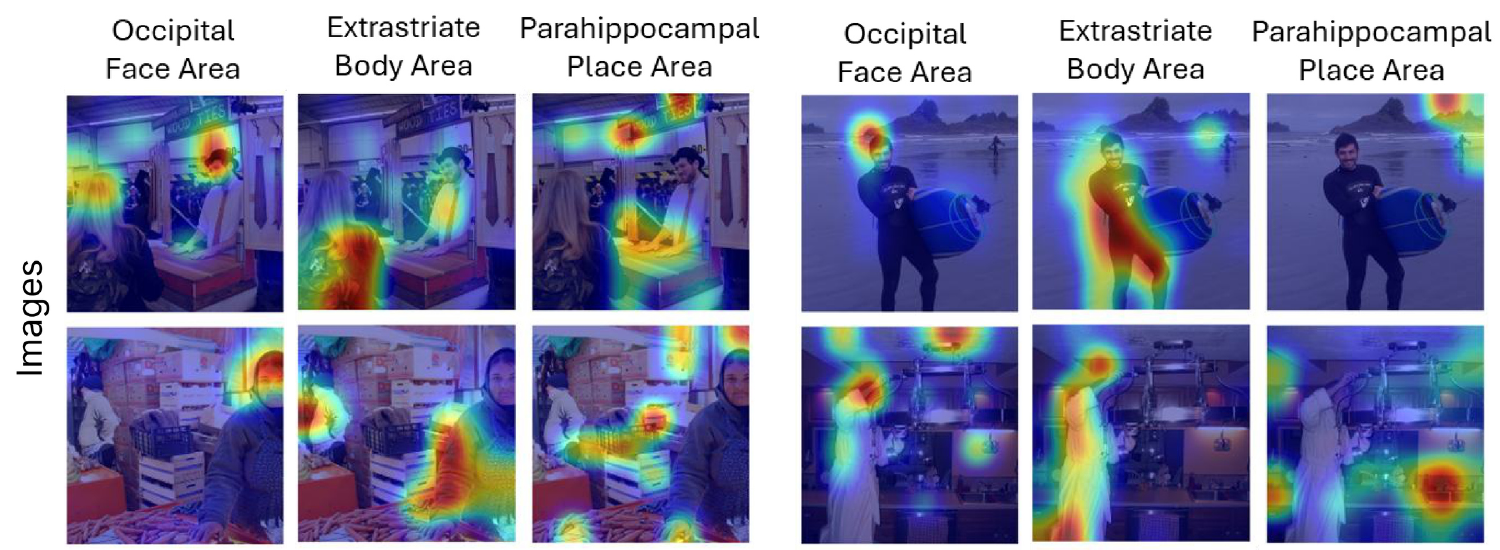}
\caption{
Spatial prototype activation maps for three category-selective regions of the visual cortex. See Appendix \ref{sup:more_spatial} for more examples. 
}
\label{fig:spatial-attention}
\end{figure}

To examine which parts of a visual scene are prioritized by each brain area, we derive spatial attention maps from the prototype activations associated with each ROI. For each ROI, we identify its top 10 prototypes using the linear prototype–voxel projection matrix. We then aggregate the spatial activations of these prototypes from the prototype representation $\mathbf{A} \in \mathbb{R}^{M \times P}$ to obtain a single $8\times8$ spatial response map. After normalization and upsampling, this produces a continuous attention map highlighting the image regions that most strongly contribute to the predicted activity of the ROI.

The resulting maps reveal distinct spatial selectivity across visual regions (Figure~\ref{fig:spatial-attention}). 
Prototypes associated with the occipital face area (OFA) selectively emphasize facial regions, whereas those linked to the extrastriate body area (EBA) concentrate on body parts and human figures. In contrast, prototypes associated with the parahippocampal place area (PPA) exhibit broader and more distributed patterns that capture scene-level structures such as room layouts. These results demonstrate that the prototype representation preserves spatial structure, enabling the model to reveal not only \emph{what} visual concepts drive neural responses but also \emph{where} those features are located within the scene.

\subsection{Prototype Manipulations Enable Controlled Stimulus Interventions}
\begin{figure}
    \centering
    \includegraphics[width=0.8\linewidth]{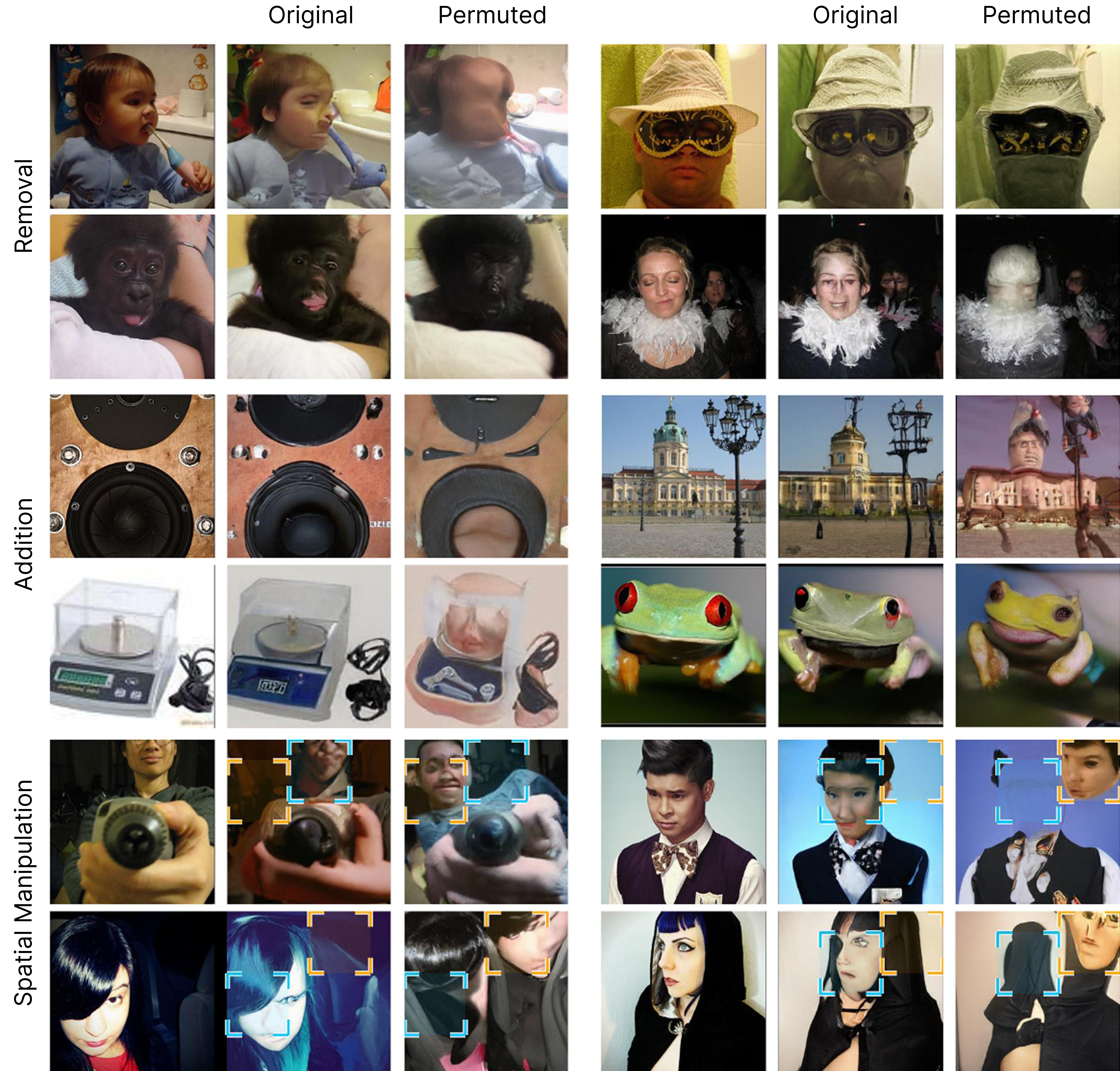}
    \caption{
    Causal interventions on face prototypes. Each example (from left to right) shows the stimulus, and its reconstruction with original/permuted prototypes. Blue and orange boxes indicate regions with high and low prototype activations that are permuted. See Appendix~\ref{sup:sup:more-causal-intervention} for more results.
    }
    \label{fig:causal_intervention}
    \vspace{-0.5em}
\end{figure}

A key advantage of our method is that individual prototypes can be selectively manipulated, enabling targeted manipulation of visual features for testing \textit{causal} hypotheses about selectivity in the brain. We demonstrate this unique capability by training a diffusion model (see Appendix \ref{sup:diffusion}) that reconstructs the original image based on its prototype activation map (\textit{i.e.,} $\mathbf{A}$ in Equation \ref{eq:proto_recon}), and selectively permuting the activation map during inference. We train our own model from scratch instead of adopting a generative model pretrained on language data (\textit{e.g.,} \cite{Rombach_2022_CVPR}), so that our analysis is less prone to potential biases \cite{brainbits}. We experiment with the top-20 prototypes for face-selective areas, and perform three types of permutation for causal intervention (see Appendix \ref{sup:sup:body-causal-intervention} for interventions with body prototypes): 

\paragraph{Removal.} 
We choose images that strongly activate the selected prototypes, and set those prototype activations to zero before generating images from the modified representation. The resulting images (1st-2nd row in Figure \ref{fig:causal_intervention}) show faces selectively removed from the scene while other objects and background elements remain largely intact, indicating that the face prototype captures face-specific information without strongly entangling unrelated visual content.

\paragraph{Addition.}
Conversely, increasing the activation of face prototypes by their mean values induces the appearance of face-like features in previously non-face stimuli, yielding pareidolia-like objects with visually salient eye- and mouth-like structures (3rd–4th row in Figure \ref{fig:causal_intervention}).
\begin{wrapfigure}{r}{0.3\textwidth}
    \vspace{-1em}
    \setlength{\columnsep}{8pt}
    \centering
    \includegraphics[width=\linewidth]{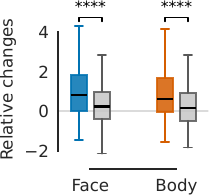}
    \caption{Changes in predicted fMRI responses for targeted permutations. Gray boxes denote controls.}
    \label{fig:close-loop}
    \vspace{-1.7em}
\end{wrapfigure}
In face stimuli, the same manipulation further magnifies these features, for example, by enlarging the mouth.

\paragraph{Spatial Manipulation.} 
We can also spatially manipulate the prototypes, by switching prototype activations of a neighboring region ($3\times 3$) with strong face responses (blue box) with one that has lower responses (orange box). Results in the 5th-6th row of Figure \ref{fig:causal_intervention} demonstrate that we successfully shift the locations of faces, without significantly affecting the content in unaltered regions. These interventions demonstrate that prototypes provide a causally meaningful decomposition of visual content: removing a prototype eliminates its associated features, amplifying it introduces or strengthens them, spatially manipulating them alters the corresponding structure of the scene. 

Next, we perform a \textbf{close-loop experiment} to study how targeted permutation drives brain activity. Specifically, we pass the images reconstructed with the original and permuted prototypes to our pipeline, and measure the relative changes in predicted fMRI voxel responses. We consider permutation by increasing prototype activations (\textit{i.e.,} Addition) on face and body ROIs (each with 1000 images), which is relevant to identifying the ``optimal stimuli'' for brain areas. For control groups, we perform permutation on the same set of images but with randomly selected prototype, instead of those identified based on fMRI encoding models. As visualized in Figure \ref{fig:close-loop}, targeted permutations significantly increase predicted responses in corresponding brain ROIs. On the other hand, random permutations (Control) result in much smaller changes (Average relative changes for targeted permutation vs control. Face ROIs: 0.81 vs 0.24, $p < 10^{-4}$;  Body ROIs: 0.61 vs 0.16, $p < 10^{-4}$). 

Together, the aforementioned results suggest the proposed prototype representation as a practical interface for designing stimulus manipulations to probe the selectivity of visual cortex.

\section{Conclusion, Broader Impacts, and Limitations}
This paper introduces a prototype-based framework for studying visual representations in the human brain. Our central contribution is to transform dense features from vision foundation models into sparse, spatially grounded prototypes that capture coherent visual concepts and can be linked to fMRI responses through a transparent linear mapping. This design turns an opaque visual representation into an interpretable interface between images and cortex: prototypes describe \emph{what} concepts are present and \emph{where} they occur, while voxel weights reveal \emph{which} concepts drive each brain region. Through this shared interface, we unify analyses that are typically studied separately, including neural encoding, categorical selectivity, cortical organization, spatial attribution, and controlled stimulus manipulation. More broadly, our results show how interpretable representation learning can turn foundation models from black-box predictors into structured tools for probing biological intelligence.

Our work also has room for improvement. First, we evaluate the approach in established brain areas, whereas applying it to less understood regions may enable more novel discoveries. Second, with vision foundation models trained largely with naturalistic images, our method may be less effective for interpreting synthetic or atypical stimuli. Finally, our prototypes summarize recurring visual patterns and provide an interpretable basis for neural prediction, but they do not necessarily correspond to biological prototypes implemented by the brain.

\begin{ack}
This work was supported by Howard Hughes Medical Institute (to SC and DT), and Hankuk University of Foreign Studies Research Fund (to SA).
\end{ack}

{\small
\bibliographystyle{unsrt}
\bibliography{ref}
}

%%%%%%%%%%%%%%%%%%%%%%%%%%%%%%%%%%%%%%%%%%%%%%%%%%%%%%%%%%%%
\clearpage
\appendix

\clearpage
\appendix

\section{Visualization of More Prototypes}
\label{sup:visualize-more-prototypes}

\begin{figure}[h]
    \centering
    \includegraphics[width=1\linewidth]{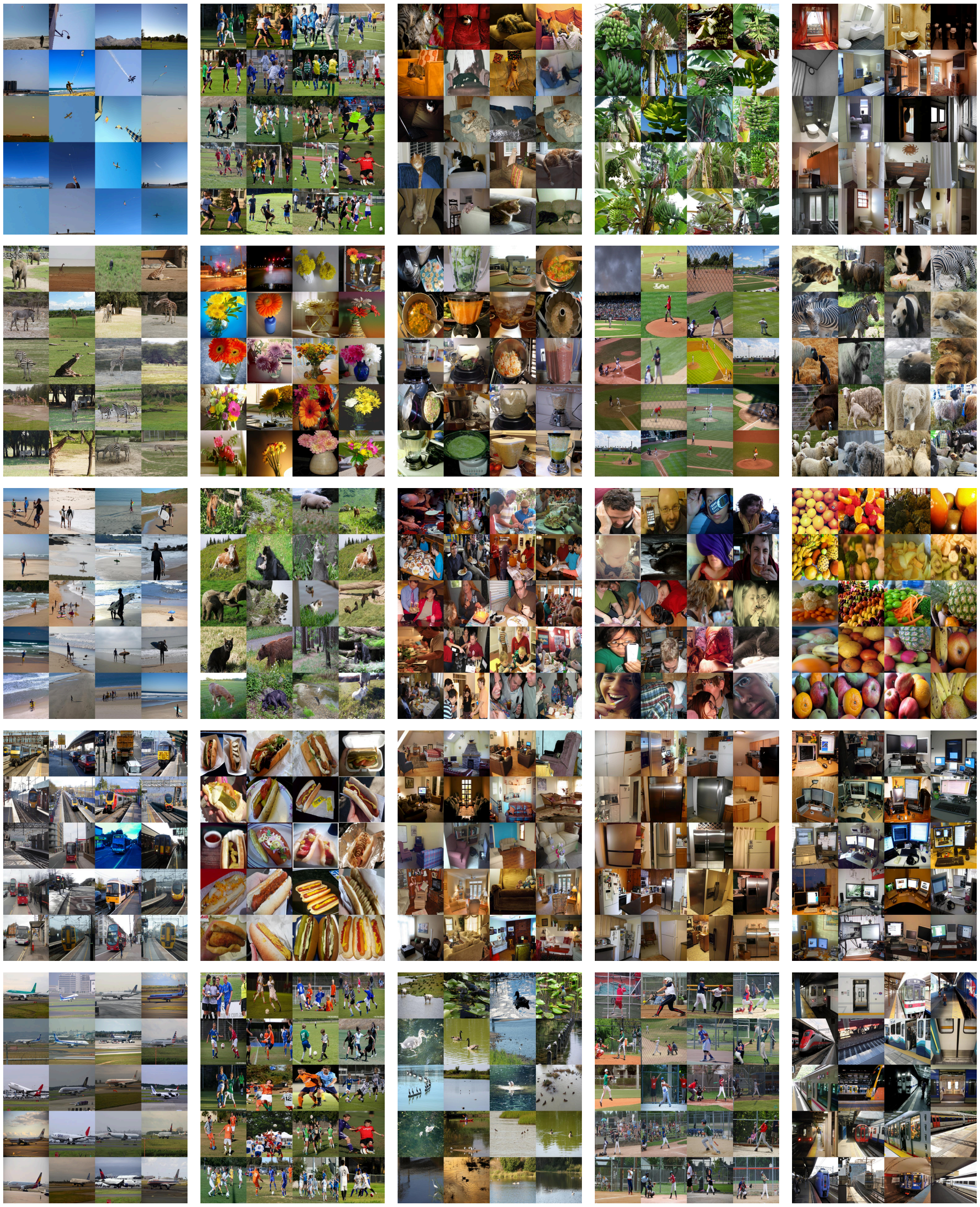}
    % \caption{Additional examples of learned prototypes. For each prototype, we show representative images from the COCO dataset that produce the highest activation scores. Consistent with the examples shown in the main text, the learned prototypes capture a diverse set of semantically coherent visual concepts, including airplanes, buildings, faces, food, sports activities, crowds, ties, electronic appliances, skiing scenes, and fruits.}
    \caption{Supplementary results for the visualization of representative images for different prototypes.}
    \label{supfig:proto_1}
\end{figure}

\begin{figure}[h]
    \centering
    \includegraphics[width=0.95\linewidth]{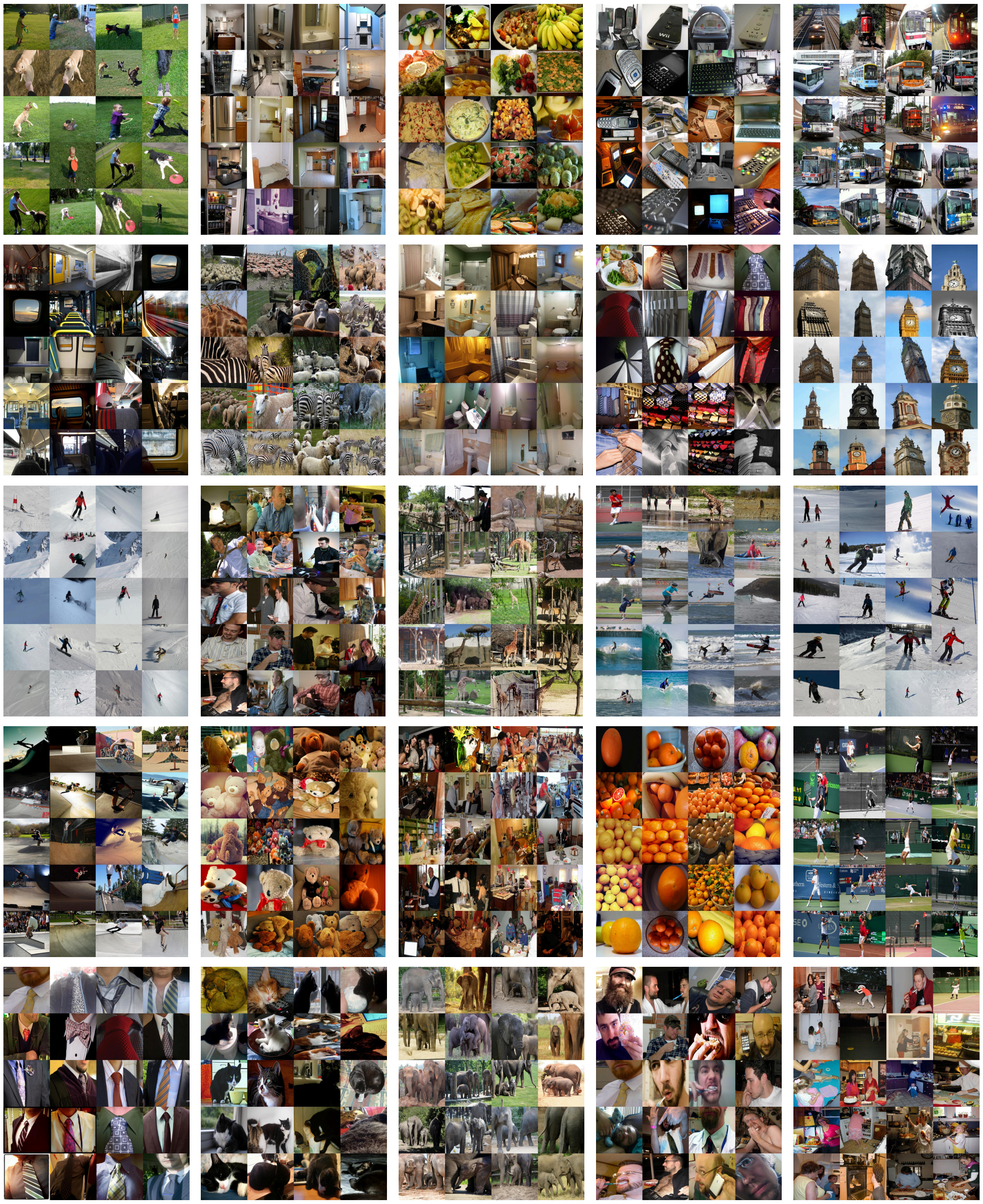}
    % \caption{Additional examples of learned prototypes. For each prototype, we show representative images from the COCO dataset that produce the highest activation scores. Consistent with the examples shown in the main text, the learned prototypes capture a diverse set of semantically coherent visual concepts, including airplanes, buildings, faces, food, sports activities, crowds, ties, electronic appliances, skiing scenes, and fruits.}
    \caption{Supplementary results for the visualization of representative images for different prototypes (cont'd).}
    \label{supfig:proto_2}
\end{figure}

Additional examples of learned prototypes (Figure \ref{supfig:proto_1} and Figure \ref{supfig:proto_2}). For each prototype, we show representative images from the COCO dataset \cite{lin2014microsoft} that produce the highest activation scores. Consistent with the examples shown in the main text, the learned prototypes capture a diverse set of semantically coherent visual concepts, including airplanes, buildings, faces, food, sports activities, crowds, ties, electronic appliances, skiing scenes, and fruits.

\clearpage
\section{Shared Prototypes Between Individuals}
\label{sup:shared-prototypes}
\begin{figure}[h]
    \centering
    \includegraphics[width=1\linewidth]{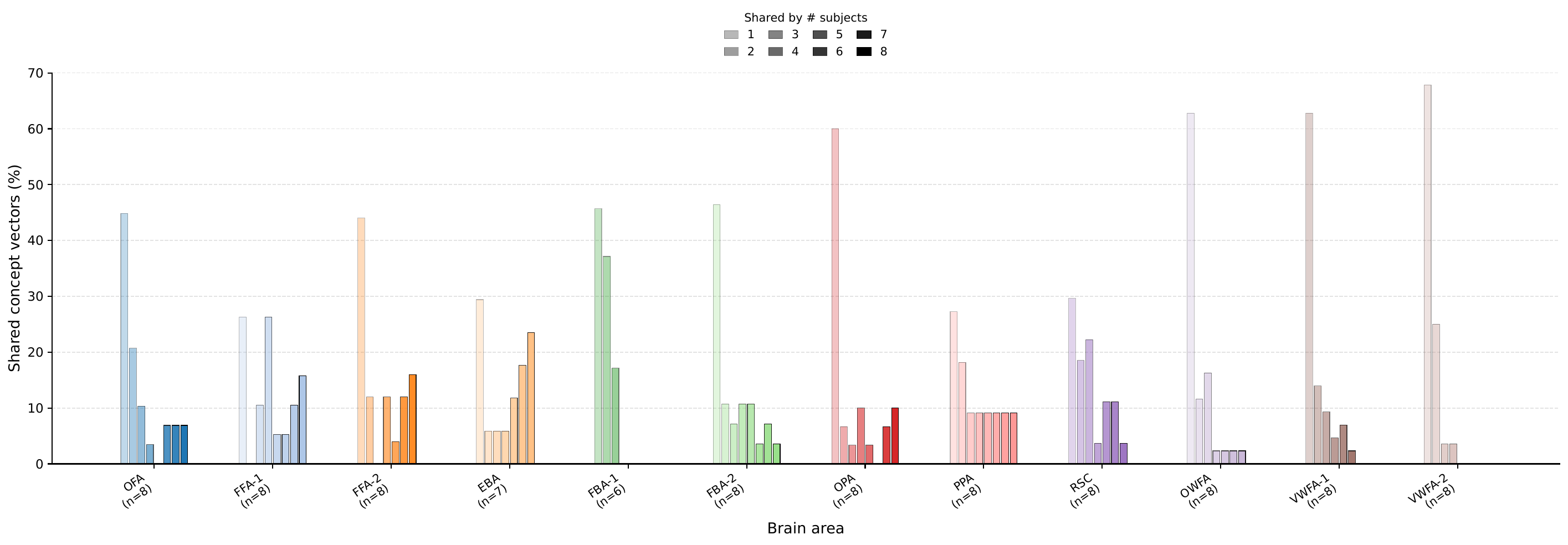}
    \caption{Histogram for the number of prototypes shared across multiple subjects.}
    \label{subfig:shared_prototype}
\end{figure}

Cross-subject consistency of selected prototypes (Figure \ref{subfig:shared_prototype}). For each subject and ROI, the top-$10$ prototypes are identified from independently trained encoding models. The histogram shows the number of prototypes shared across different numbers of subjects. Many prototypes are consistently selected across individuals, particularly for face- and body-selective regions, suggesting that these prototypes capture visual features that generalize across brains.
    
% \clearpage
\section{More ROI-Prototype Visualizations}
\label{sup:more-roi-profile}
\begin{figure}[h]
    \centering
    \includegraphics[width=0.8\linewidth]{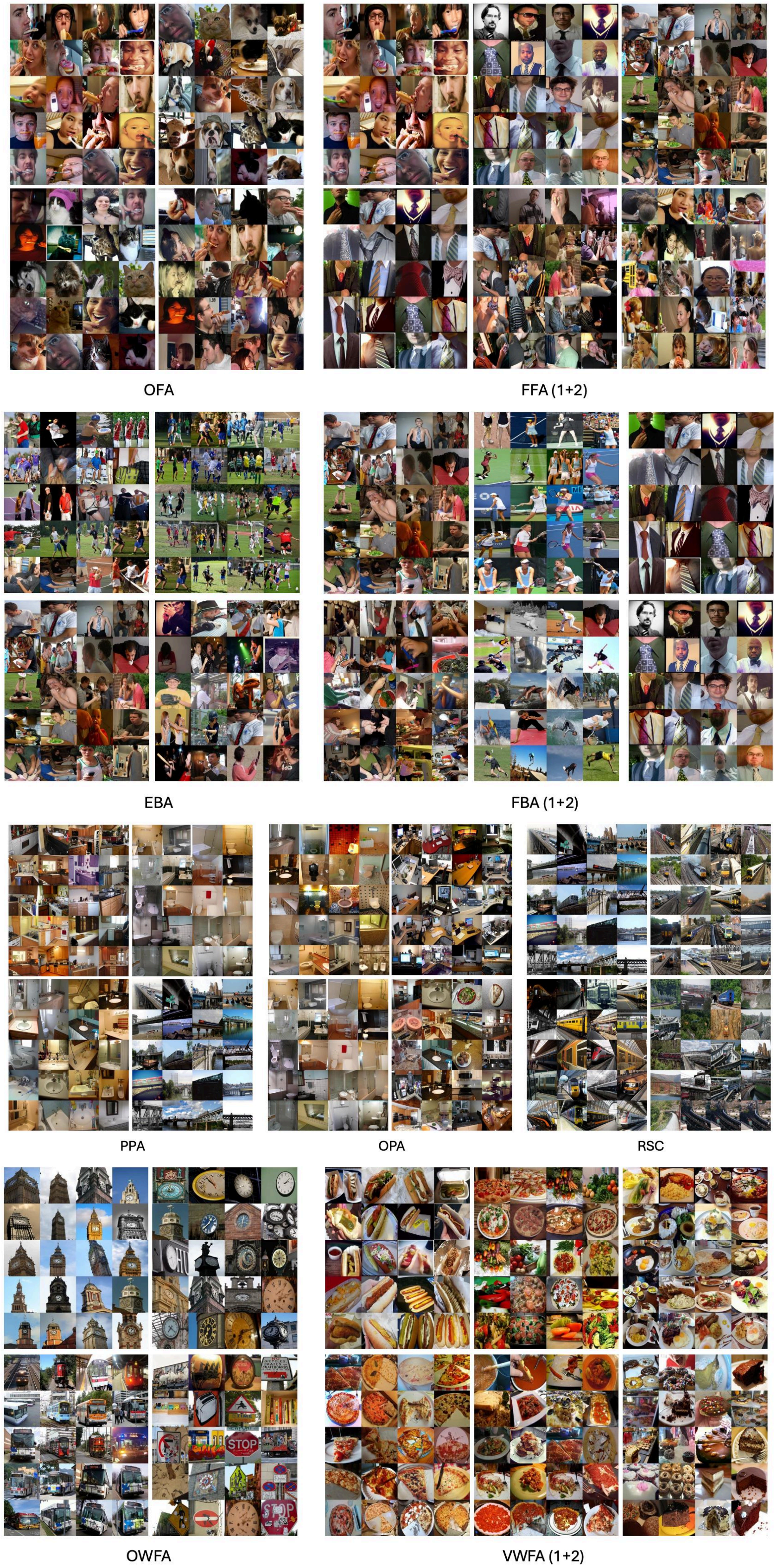}
    \caption{Top prototypes that characterize all 12 selected brain ROIs.}
    \label{supfig:roi_selectivity}
\end{figure}

More visualizations of representative prototypes for all twelve visual ROIs (Figure \ref{supfig:roi_selectivity}). The twelve visual ROIs include face-selective regions (occipital face area, OFA; fusiform face areas, FFA-1 and FFA-2), body-selective regions (extrastriate body area, EBA; fusiform body areas, FBA-1 and FBA-2), scene-selective regions (occipital place area, OPA; parahippocampal place area, PPA; retrosplenial cortex, RSC), and word-selective regions (occipital word-form area, OWFA; visual word-form areas, VWFA-1 and VWFA-2). 

These visualizations reveal additional patterns consistent with the known categorical organization of the ventral visual stream. For example, the occipital word-form area (OWFA) is driven by prototypes containing objects with prominent lettering (e.g., clocks, stop signs, and bus fronts with visible text). The visual word-form area (VWFA) unexpectedly shows a prevalence of food-related prototypes. One possible explanation is that these images share visual properties, such as dense textures, repeated circular elements, or fine-grained patterns that resemble the spatial statistics of letter strings, which are known to drive responses in this region.

\clearpage
\section{Visualization of Prototypes with Diverse Weights to ROIs}
\label{sup:roi-profile-spectrum}
Our results in the main paper and Appendix \ref{sup:more-roi-profile} visualize the most preferred prototypes for individual ROIs, demonstrating the ability of our method to capture the dominant selectivity of brain ROIs. In this section, we present complementary results showing the full spectrum of prototype preferences—ranging from the most to the least preferred prototypes for predicting responses in a given ROI.
\begin{figure}[h]
    \centering
    \includegraphics[width=1\linewidth]{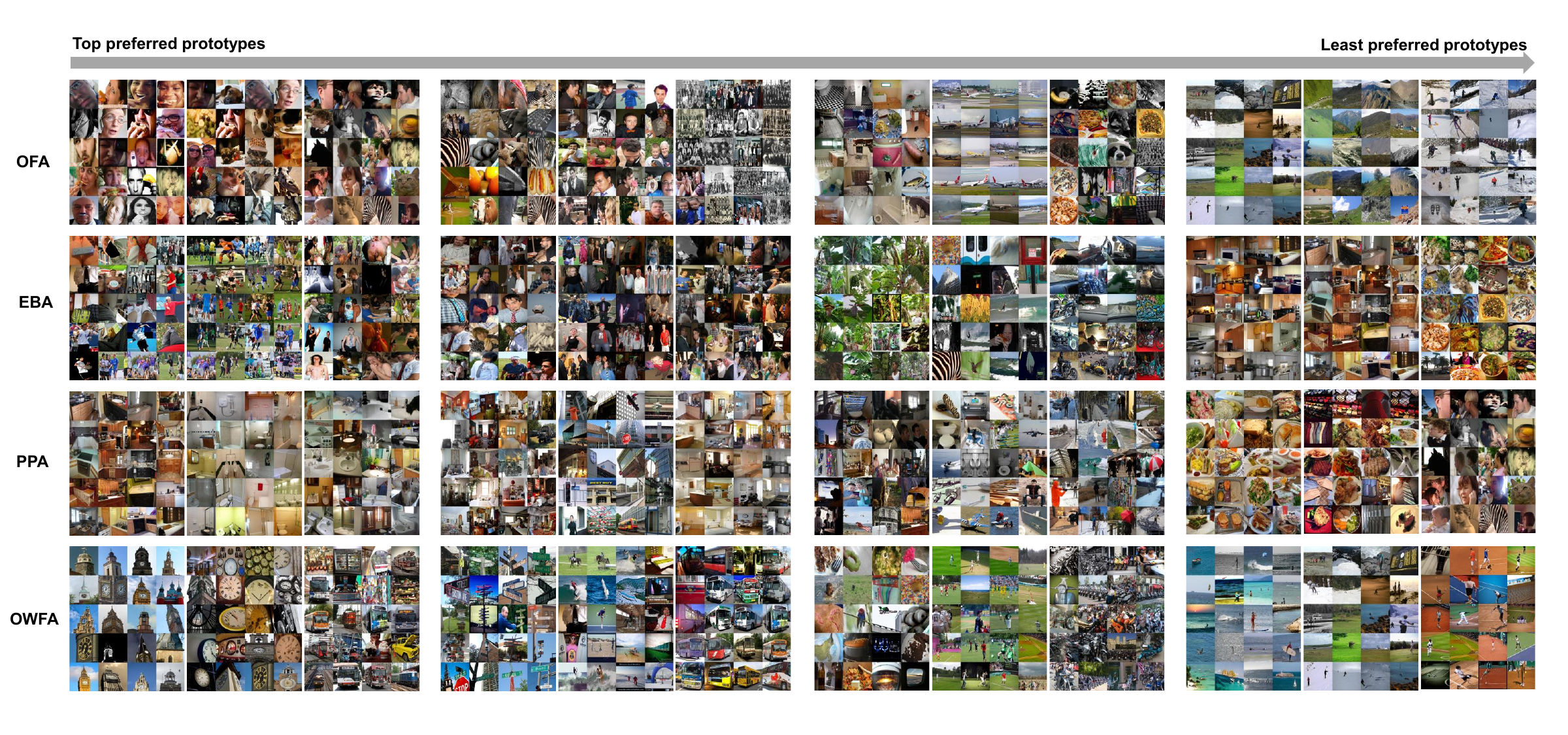}
    \caption{Prototypes with different preference levels for four brain ROIs with distinct selectivity.}
    \label{supfig:proto_spectrum}
\end{figure}

As shown in Figure \ref{supfig:proto_spectrum}, our method further characterizes the continuous tuning properties of brain ROIs. For the occipital face area (OFA), moving from the most to the least preferred prototypes reveals a gradual transition from frontal faces, to human figures in varied poses, and finally to natural scenes. A similar trend holds for the extrastriate body area (EBA), where prototypes shift from sports scenes with prominent body movements, to group photos, and eventually to indoor scenes without people. Together, these results show that our method characterizes brain ROIs beyond their dominant selectivity, revealing a graded preference structure that reflects their tuning.

\section{Effects of Spectral Clustering on Prototype Visualization}
\label{sup:clustering_effect}
\begin{figure}[h]
    \centering
    \includegraphics[width=1\linewidth]{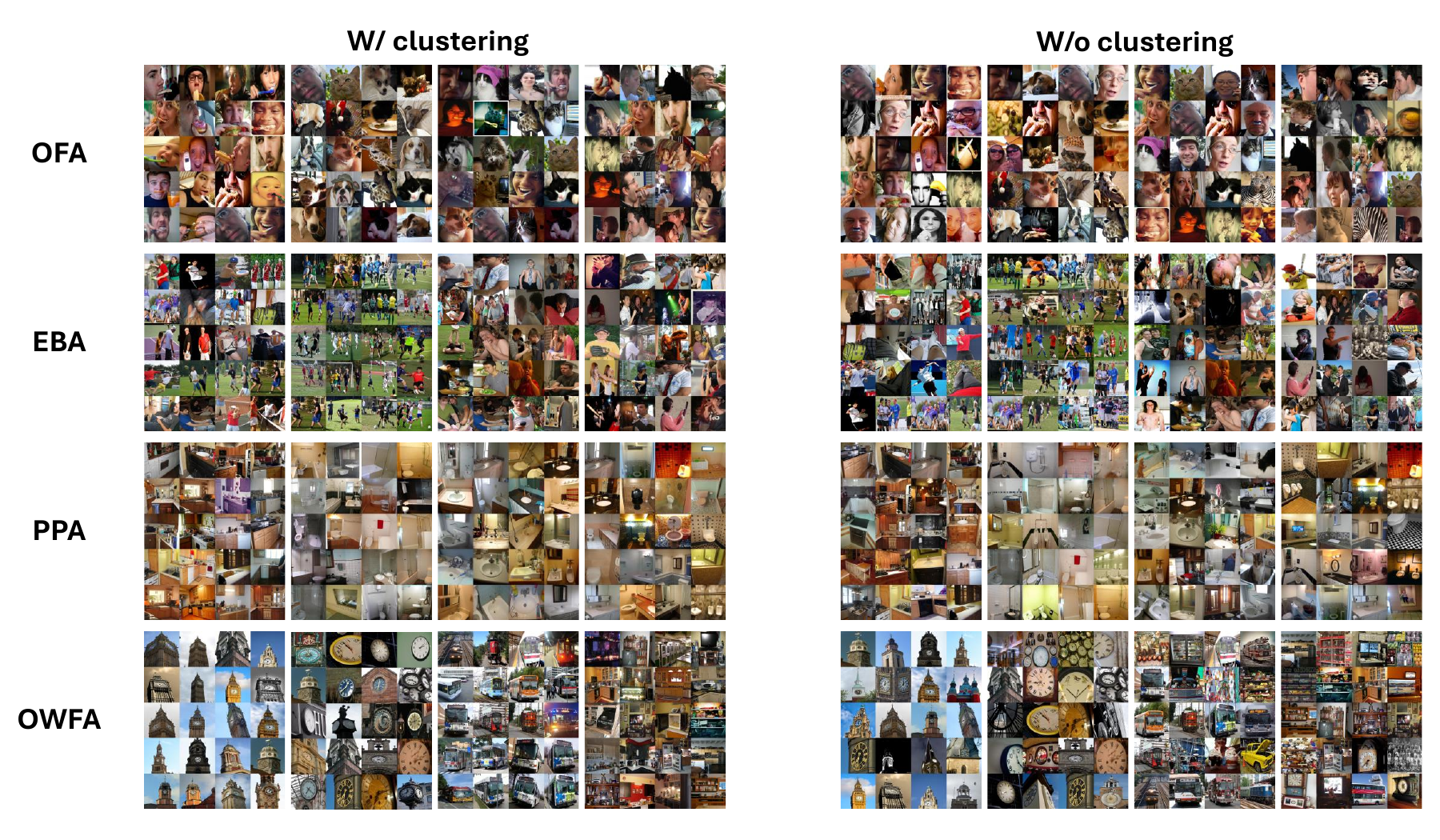}
    \caption{Representative images for the same prototypes (\textit{i.e.,} Top-4 prototypes for different brain ROIs) with and without clustering.}
    \label{supfig:proto_clustering}
\end{figure}

Following \cite{brainexplore} that uses CLIP embedding for image selection and pattern summarization, we employ spectral clustering with CLIP embedding to select representative images for each prototype. This step is optional and for visualization only. To evaluate its influence, we compare the selected images for the same prototypes before and after clustering in Figure \ref{supfig:proto_clustering}. Results show that, while clustering affects selection of specific images, it does not change the semantic meanings of the corresponding prototypes, \textit{e.g.,} frontal faces for OFA, sports scenes for EBA, and texts for OWFA.

\section{PCA Fails to Characterize Brain ROIs}
\label{sup:pca-roi-profile}
\begin{figure}[h]
    \centering
    \includegraphics[width=1.0\linewidth]{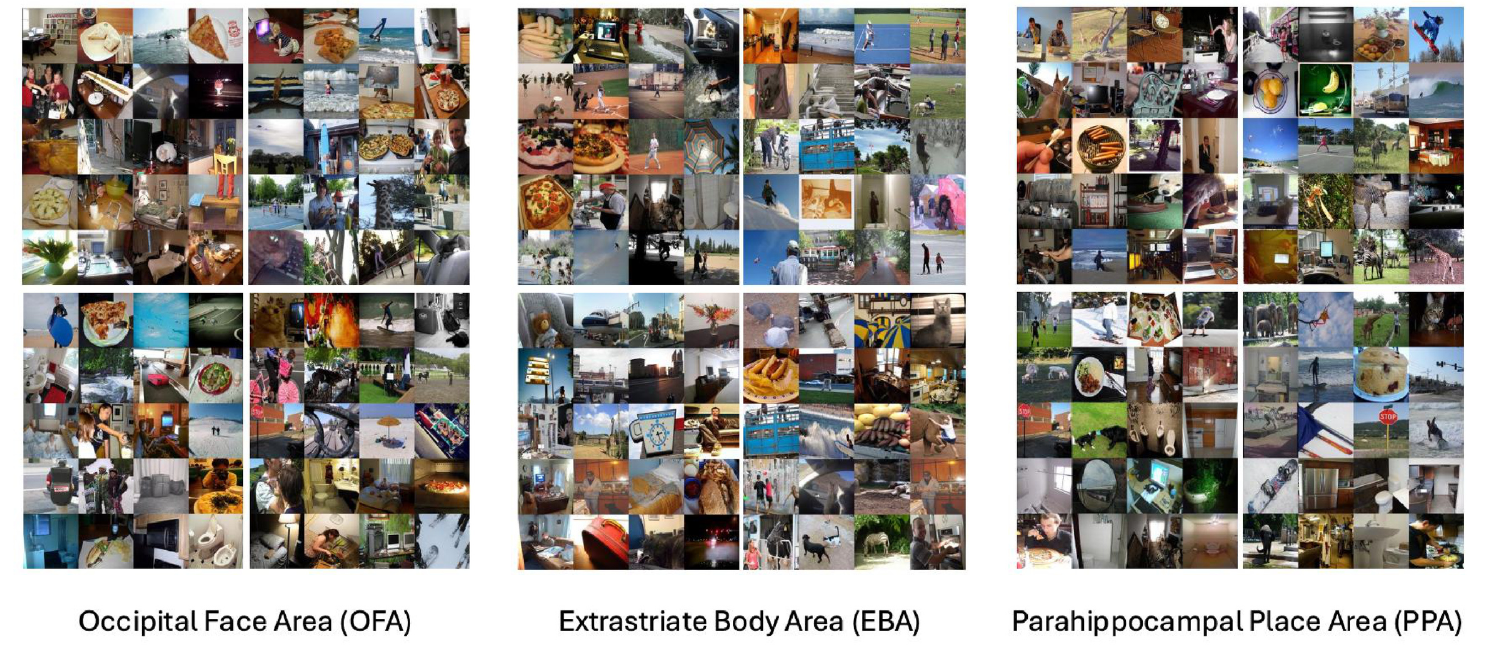}
    \caption{Representative images of top principal components (for PCA features) that characterize different brain ROIs.}
    \label{supfig:pca_selectivity}
    
\end{figure}

As a control baseline for prototypes, we apply Principal Component Analysis (PCA) on the raw DINOv3 \cite{dinov3} features, using the same training data as our method. We then train a fMRI encoding model based on the PCA features, and identify the top principal components (PCs) \cite{pca} for different brain areas. In Figure \ref{supfig:pca_selectivity}, we visualize top-20 images with the highest values for each of these selected PCs (top-4 PCs shown for each brain ROI). The results show that, unlike our visual prototypes, PCA features do not exhibit clear category-selective patterns across visual regions. In addition, the top images tuned for each PC also do not have coherent meanings.

% ROI feature visualizations using PCA-reduced DINO features. Representative images associated with the principal components that most strongly contribute to each ROI are shown. Unlike the prototype-based analysis, PCA features do not exhibit clear category-selective patterns across visual regions.
    
\section{Comparison between Prototypes and Sparse Autoencoder for Brain ROI Characterization}
\label{sup:sae_proto_comparison}
In the main paper, we show that our proposed prototype method achieves better fMRI encoding performance compared to a sparse autoencoder (SAE) learned on top of DINOv3 features. In this section, we further demonstrate the advantages of our method in characterizing the selectivity of brain ROIS. In particular, we visualize the representative images for the top-5 prototypes (our method) or SAE features (SAE) with the highest weights for predicting fMRI responses in different brain ROIs (Occipital Face Area, Extrastriate Body Area, and Parahippocampal Place Area). 

\begin{figure}[h]
    \centering

    \begin{subfigure}{0.9\linewidth}
        \centering
        \includegraphics[width=\linewidth]{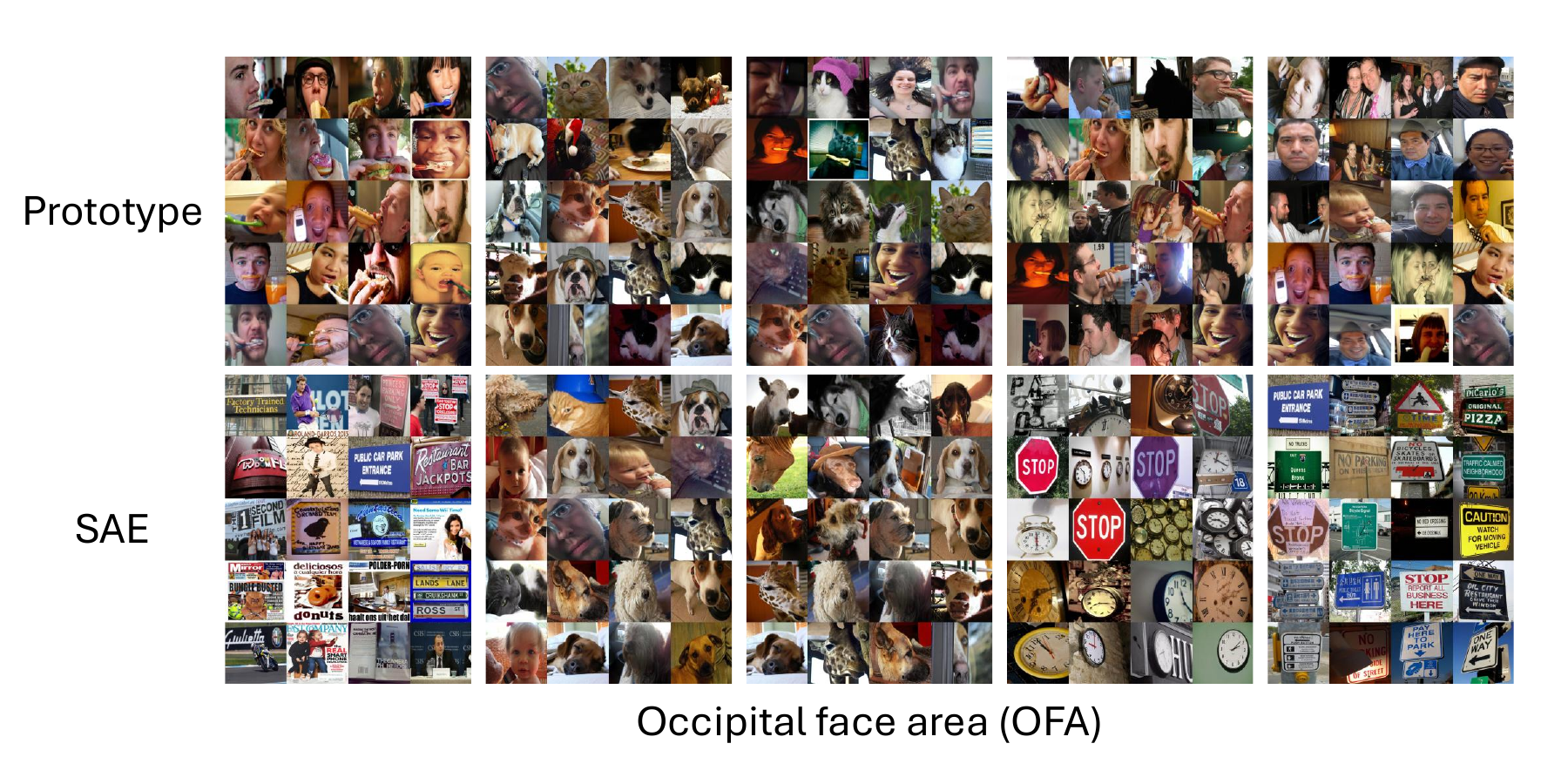}
    \end{subfigure}

    \vspace{0.5em}

    \begin{subfigure}{0.9\linewidth}
        \centering
        \includegraphics[width=\linewidth]{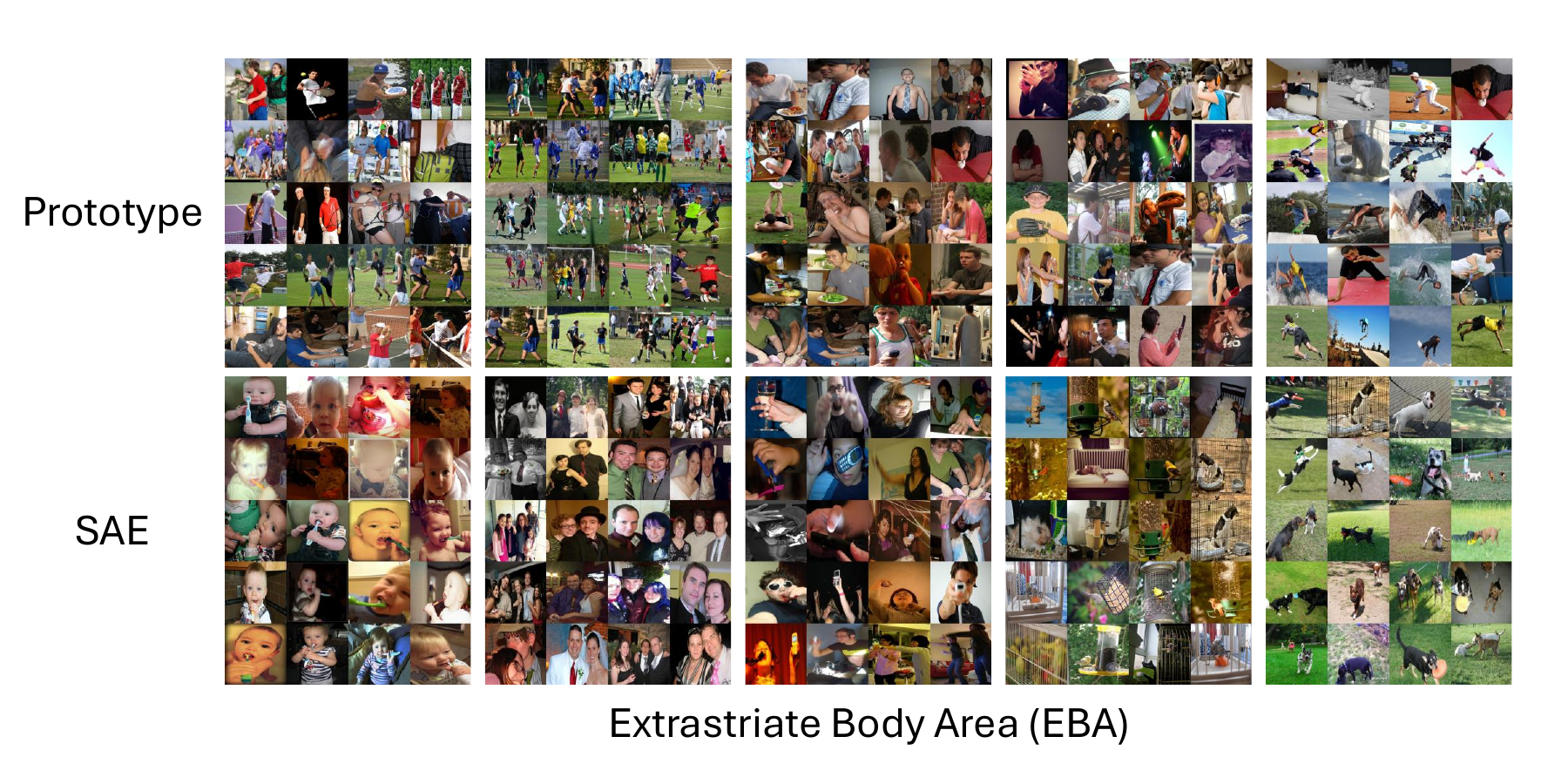}
    \end{subfigure}

    \vspace{0.5em}

    \begin{subfigure}{0.9\linewidth}
        \centering
        \includegraphics[width=\linewidth]{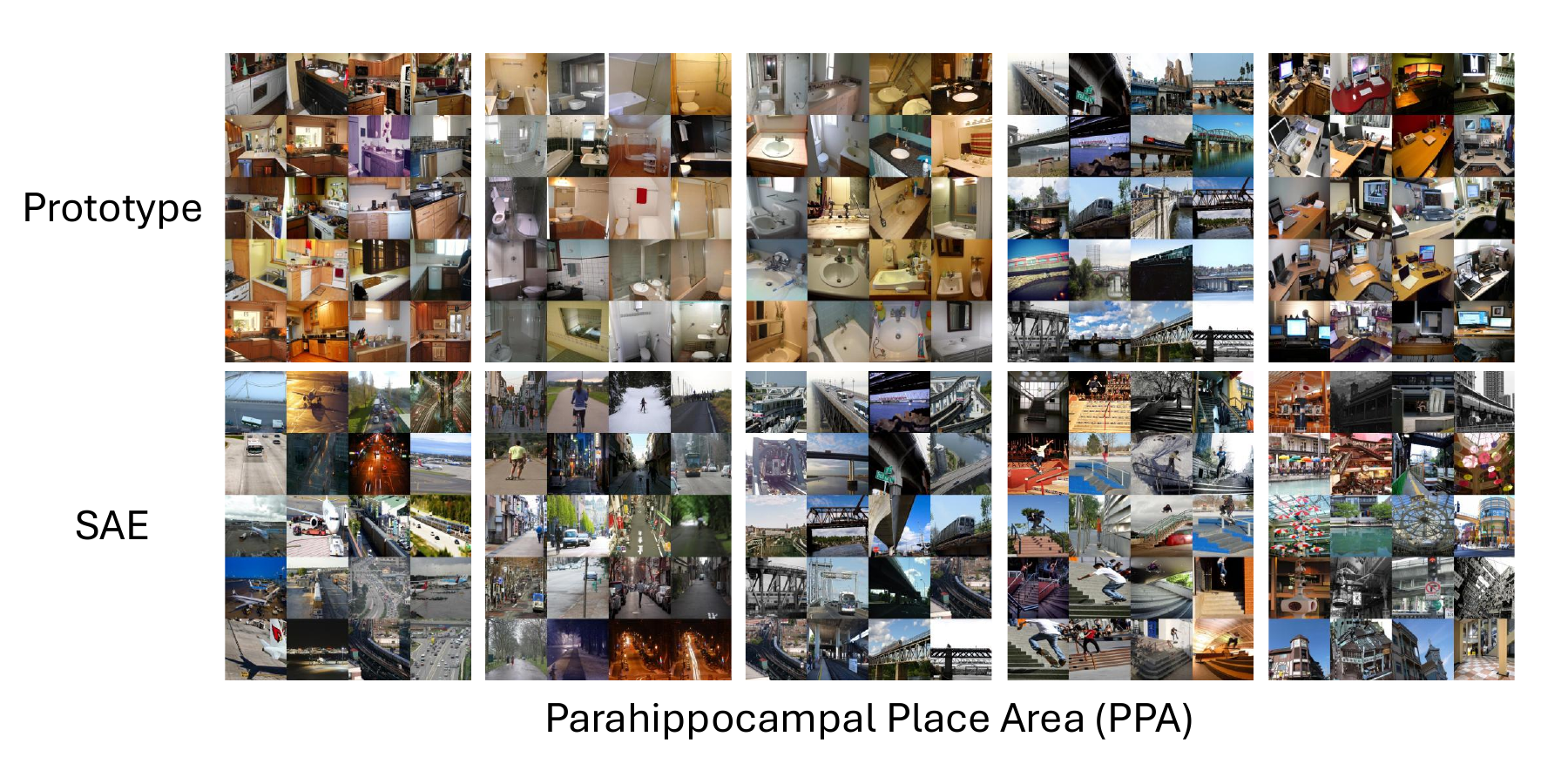}
    \end{subfigure}
    \caption{Comparison between the proposed prototype method (Prototype) and sparse autoencoder (SAE) for characterizing selectivity of different brain ROIs. Each row visualizes the representative images for top-5 prototypes/SAE features.}
    \label{supfig:sae_proto_comparison}
\end{figure}

As shown the in Figure \ref{supfig:sae_proto_comparison}, SAE roughly captures the selectivity of certain brain ROIs (\textit{e.g.,} top SAE features for Parahippocampal Place Area mostly correspond to outdoor scenes). However, unlike our method that characterize a brain ROI with prototypes with coherent meanings (\textit{e.g.,} close-up faces for Occipital Face Area and sports scenes for Extrastriate Body Area), top SAE features for a brain ROI typically have mixed semantics (\textit{e.g.,} top SAE features for Occipital Face Area are associated to both text and faces, top SAE features for Extrastriate Body Area show close-up faces or animals with limited movement), resulting in difficulties in accurately elaborating the brain selectivity. 

These results demonstrate the effectiveness of our approach. More importantly, they highlight the advantages of explicitly factorizing visual features into compositional prototypes compared to learning implicit sparse features.

\clearpage

\section{Neural Encoding Performance of Top Preferred Prototypes}
\label{sup:prototype_predictivity}
Our analysis in the main paper show that prototypes with the largest positive weights successfully reveal an ROI’s dominant preferences (\textit{e.g.,} face, body, scene), and the full collection of prototypes show improved fMRI encoding performance compared to the corresponding baselines. In this section, we ask a complimentary question: how much predictive performance is retained when the encoding model is restricted to its top-(k) preferred prototypes? We address this question by ranking prototype features by their learned encoding weights and fitting a separate linear regression model for each voxel using only the top-(k) features.
\begin{table}[h]
\centering
\resizebox{\linewidth}{!}{
\begin{tabular}{lccccccccccccc}
\hline
 & OFA & FFA-1 & FFA-2 & EBA & FBA-1 & FBA-2 & OPA & PPA & RSC & OWFA & VWFA-1 & VWFA-2 & Average \\
\hline
Top-4 & 0.184 & 0.283 & 0.358 & 0.295 & 0.183 & 0.327 & 0.118 & 0.126 & 0.262 & 0.100 & 0.161 & 0.169 & 0.214\\
Top-8 & 0.192 & 0.318 & 0.375 & 0.313 & 0.242 & 0.329 & 0.157 & 0.197 & 0.333 & 0.153 & 0.154 & 0.213 & 0.248\\
Top-16 & 0.225 & 0.332 & 0.393 & 0.333 & 0.256 & 0.359 & 0.155 & 0.206 & 0.334 & 0.187 & 0.258 & 0.228 & 0.272\\
Top-32 & 0.274 & 0.342 & 0.411 & 0.347 & 0.263 & 0.372 & 0.180 & 0.286 & 0.342 & 0.225 & 0.261 & 0.261 & 0.297\\
Top-64 & 0.304 & 0.375 & 0.427 & 0.365 & 0.292 & 0.394 & 0.251 & 0.327 & 0.365 & 0.281 & 0.281 & 0.279 & 0.329\\
Full model & 0.371 & 0.466 & 0.516 & 0.506 & 0.403 & 0.504 & 0.414 & 0.424 & 0.414 & 0.299 & 0.390 & 0.293 & 0.416 \\
\hline
\end{tabular}
}
\vspace{0.5em}
\caption{fMRI encoding with top-k preferred prototypes, results are reported on subject 1.}
\label{tab:top_proto_fmri}
\end{table}

The results in Table \ref{tab:top_proto_fmri} show that the top-4 prototypes capture a considerable component of voxel responses, achieving a mean correlation of (r=0.214), or approximately 52$\%$ of the full-model correlation. Performance increases progressively as more prototypes are included: the top 16, 32, and 64 retain approximately $65\%$, $71\%$, and $79\%$ of the full-model correlation (with all 1000 prototypes), respectively. Together, these results indicate that a relatively small subset of the highest-weighted prototypes accounts for most of the ROI-level prediction, although the full set is needed to achieve the best predictive performance.

\section{Additional Examples for Spatial Selectivity of Brain ROIs}
\label{sup:more_spatial}
\begin{figure}[h]
    \centering
    \includegraphics[width=0.8\linewidth]{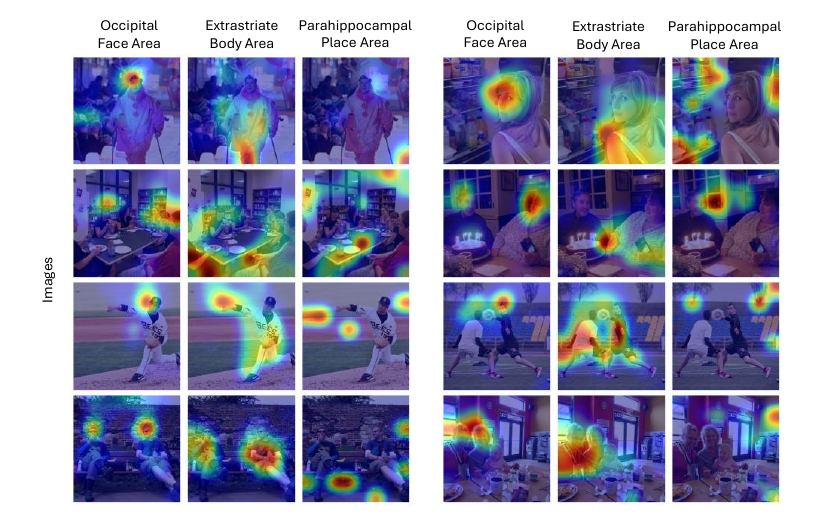}
    \caption{Additional examples for spatial selectivity of brain ROIs.}
    \label{supfig:spatial_roi}
\end{figure}

In Figure \ref{supfig:spatial_roi}, we provide additional examples for characterizing the spatial selectivity of three distinct brain ROIs (Occipital Face Area, Extrastriate Body Area, and Parahippocampal Place Area). The results further demonstrate that our approach accurately reveals the spatial locations that tightly aligned with the functional selectivity of brain ROIs.

\section{More per-prototype cortical activation maps}
\label{sup:more-proto-flatmap}

\begin{figure}[h]
    \centering
    \includegraphics[width=0.9\linewidth]{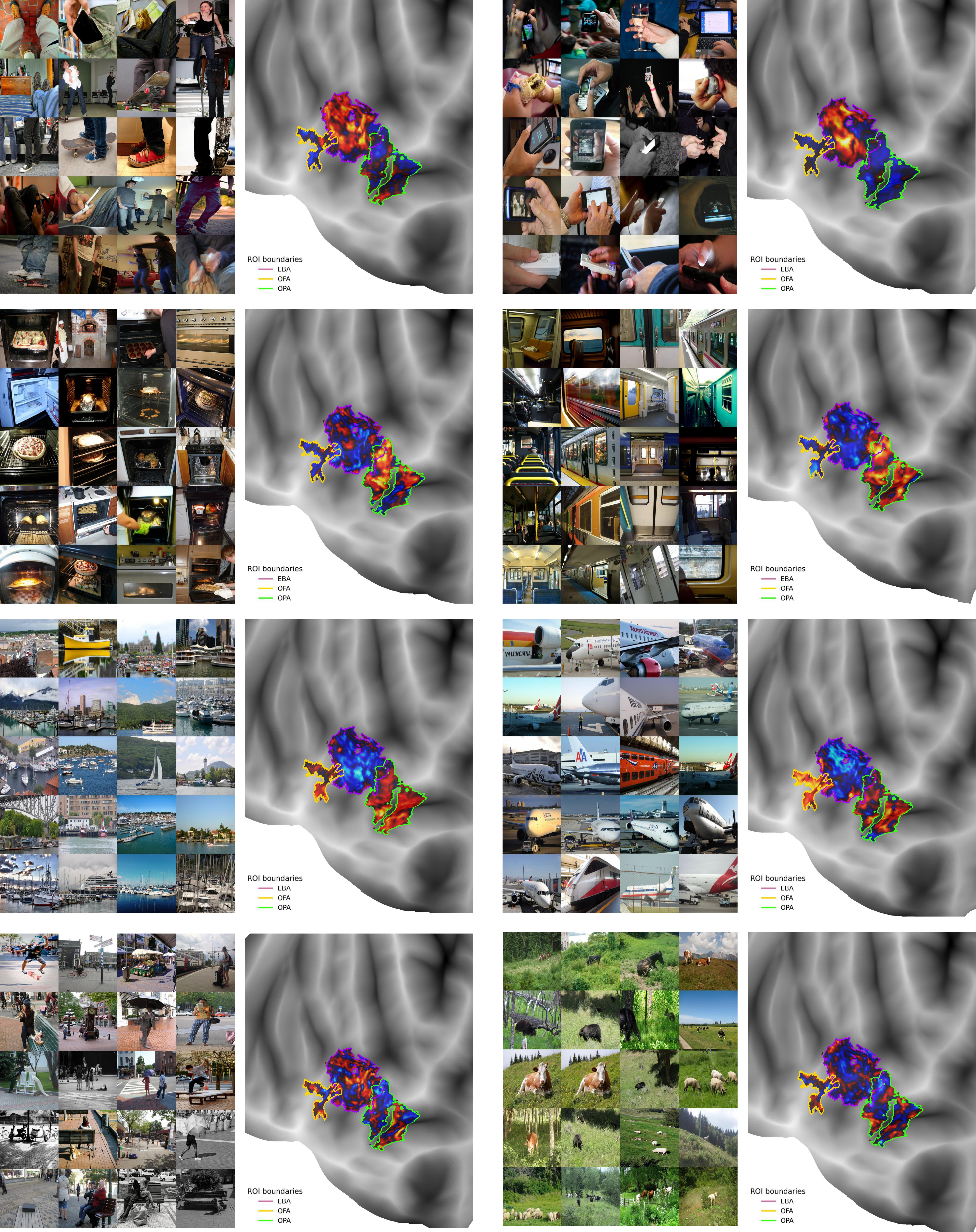}
    \caption{Additional results on prototype weights for different brain ROIs.}
    \label{supfig:proto_weight}
\end{figure}

Per-prototype cortical activation maps and representative image examples (Figure \ref{supfig:proto_weight}). For each prototype, we visualize the top activating images (left) and the corresponding cortical activation pattern across ROIs (right). Warm colors indicate stronger predicted responses, while cool colors indicate weaker responses. ROI boundaries for EBA, OFA, and OPA are outlined. 

Prototypes in the first row selectively activate the extrastriate body area (EBA), corresponding to images containing body parts such as legs, feet, and hands interacting with objects (e.g., shoes, phones). 
The second row shows prototypes associated with the occipital place area (OPA), capturing indoor environments such as kitchens and transportation scenes. 
The third row highlights prototypes that jointly activate OFA and OPA, often corresponding to scenes containing a single prominent object, e.g., airplanes or trains embedded within a scene. The final row illustrates prototypes that produce distributed activation across multiple ROIs, corresponding to complex scene categories such as outdoor environments or farm scenes. 

% \clearpage
\section{Controlled Stimulus Interventions with Body Prototypes}
\label{sup:sup:body-causal-intervention}
\begin{figure}[h]
    \centering
    \includegraphics[width=1\linewidth]{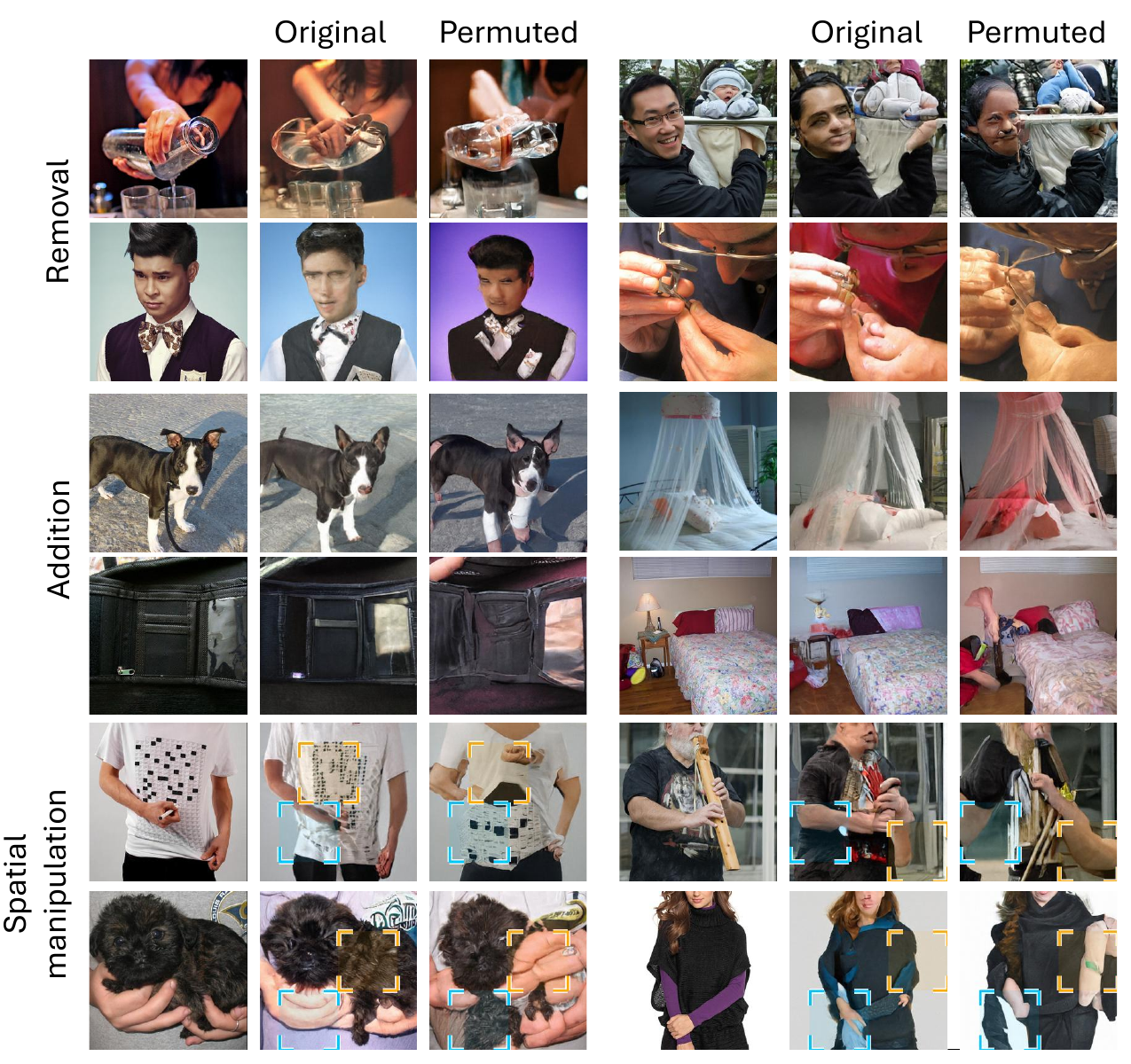}
    \caption{Causal stimuli generated results with targeted permutations on body areas.}
    \label{supfig:body_ablation}
\end{figure}

The main paper uses prototypes contributing to face-selective brain ROIs for causal intervention with our approach. In this section, we additionally experiment with top-20 prototypes that have the highest weights for body-selective ROIs, and perform the same three types of permutations. As visualized in Figure \ref{supfig:body_ablation}, manipulating these body prototypes robustly alters the targeted visual concepts (\textit{e.g.,} hands and arms) while keeping the other image properties relatively unchanged. Specifically: (1) Setting activations of body prototypes (\textit{i.e.,} Removal) to zeros removes hands and arms within the images, or distorts the shape of fingers; (2) On the other hand, increasing the activations (\textit{i.e.,} Addition) leads to emergence of hands and fingers, or alters the colors of objects to make them more skin-like; (3) Spatially permuting (\textit{i.e.,} Spatial manipulation) the body prototypes (blue/orange boxes denote regions with high/low body prototype activations that are switched) allows us to shift the locations of corresponding concepts. Taken together, these results complement our findings in the main paper, and further validate the usefulness of our approach for probing the causal relationships between neural activity and visual concepts.
\clearpage

% \clearpage
\section{Additional Examples of Face Prototype Interventions}
\label{sup:sup:more-causal-intervention}
\begin{figure}[h]
    \centering
    \includegraphics[width=0.65\linewidth]{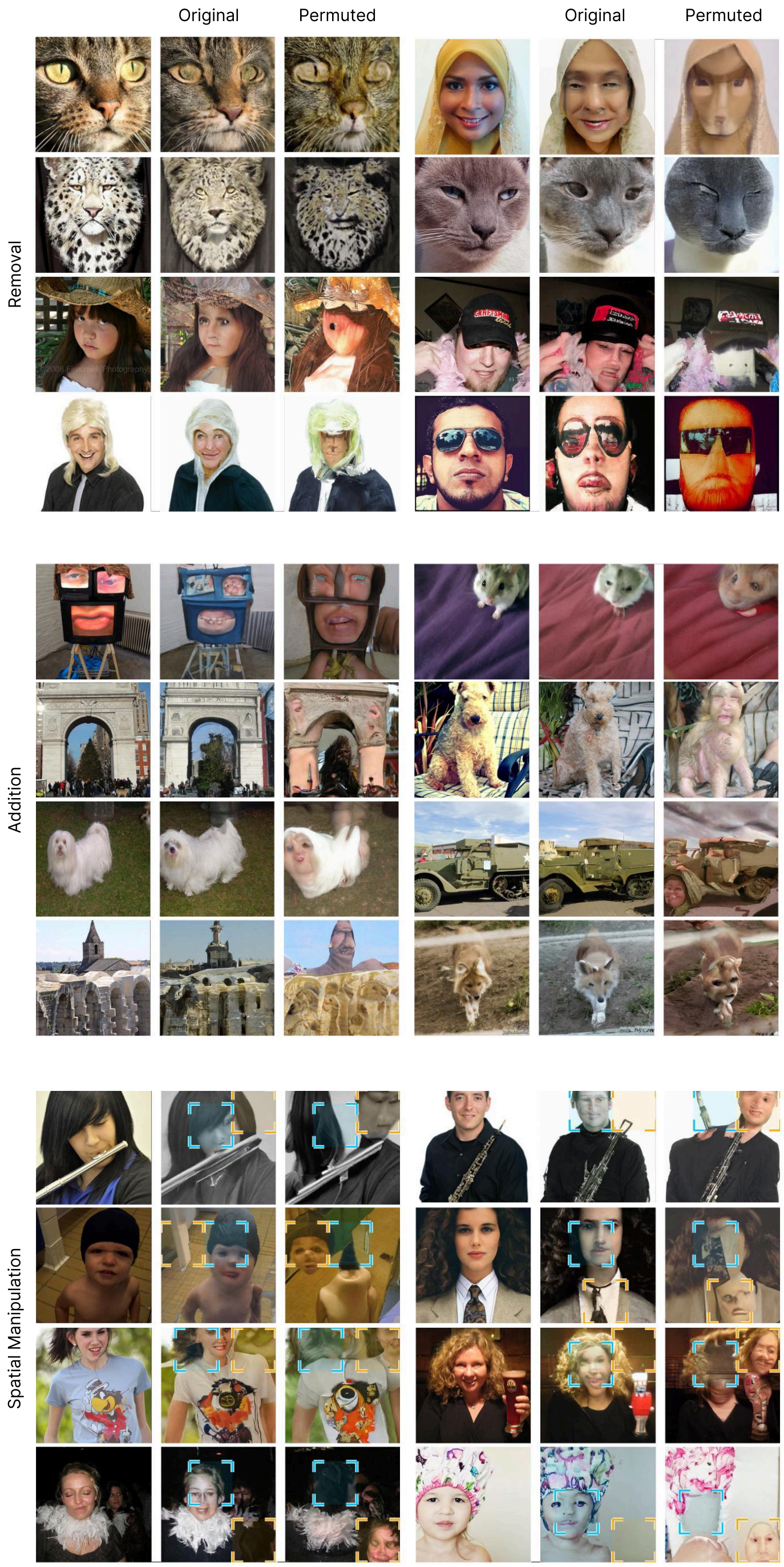}
    \caption{Additional results on causal stimuli generation with permutations of body areas.}
    \label{supfig:face_ablation}
\end{figure}

In this section, we provide additional examples (Figure \ref{supfig:face_ablation}) for causal intervention with the three types of permutations for face prototypes. The results agree with our observations in the main paper.
    
% \begin{figure}[h]
%     \centering
%     \includegraphics[width=1.0\linewidth]{figures/sup_removal.pdf}
%     % \caption{Enter Caption}
% \end{figure}
% \begin{figure}[h]
%     \centering
%     \includegraphics[width=1.0\linewidth]{figures/sup_addition.pdf}
%     % \caption{Enter Caption}
% \end{figure}
% \begin{figure}[h]
%     \centering
%     \includegraphics[width=1.0\linewidth]{figures/sup_spatial_manipulation.pdf}
%     % \caption{Enter Caption}
% \end{figure}

%%%%%%%%%%%%%%%%%%%%%%%%%%%%%%%%%%%%%%%%%%%%%%%%%%%%%%%%%%%%

\clearpage
\section{Supplementary Neural Encoding Results}
\label{sup_encoding_sec}
Table \ref{tab:roi_correlation_sup} reports the fMRI encoding performance of different models, across all 12 brain ROIs used in our experiments. 
\begin{table}[h]
\centering
\resizebox{\linewidth}{!}{
\begin{tabular}{lcccccccccccc}
\hline
 & OFA & FFA-1 & FFA-2 & EBA & FBA-1 & FBA-2 & OPA & PPA & RSC & OWFA & VWFA-1 & VWFA-2 \\
\hline
Noise-ceiling  & 0.561 & 0.597 & 0.541 & 0.600 & 0.541 & 0.588 & 0.563 & 0.585 & 0.542 & 0.530 & 0.551 & 0.500 \\
Adeli \textit{et al.} \cite{adeli2025transformer} & 0.430 & 0.489 & 0.428 & 0.481 & 0.432 & 0.476 & 0.438 & 0.485 & 0.444 & 0.399 & 0.430 & 0.363\\
\hline
JEPA-VIT & 0.305 & 0.381 & 0.324 & 0.381 & 0.348 & 0.381 & 0.348 & 0.414 & 0.368 & 0.244 & 0.317 & 0.259\\
Proto JEPA-VIT & 0.355 & 0.403 & 0.346 & 0.404 & 0.359 & 0.400 & 0.399 & 0.419 & 0.353 & 0.314 & 0.350 & 0.270\\
\hline
DINOv3-ConvNeXt & 0.273 & 0.366 & 0.308 & 0.379 & 0.319 & 0.369 & 0.324 & 0.377 & 0.330 & 0.209 & 0.304 & 0.256\\
Proto DINOv3-ConvNeXt & 0.335 & 0.410 & 0.340 & 0.421 & 0.347 & 0.404 & 0.378 & 0.408 & 0.310 & 0.268 & 0.342 & 0.267\\
\hline
DINOv3-VIT & 0.302 & 0.388 & 0.326 & 0.396 & 0.330 & 0.379 & 0.332 & 0.386 & 0.325 & 0.254 & 0.314 & 0.255 \\
Proto DINOv3-VIT & 0.367 & 0.434 & 0.376 & 0.433 & 0.387 & 0.432 & 0.404 & 0.443 & 0.387 & 0.310 & 0.370 & 0.296\\
\hline
\end{tabular}
}
\vspace{0.5em}
\caption{Comparison of fMRI encoding across 8 subjects. We report the voxel-wise Pearson's correlation for all 12 brain ROIs.}
\label{tab:roi_correlation_sup}
\end{table}

\section{Ablation Study on the Hyperparameters}
\label{sup:hyperparameter}
Our method has two hyperparameters, \textit{i.e.,} the number of prototypes $P$ and $k$ for sparsity constraint (only top-$k$ prototypes are activated). Here we show a comprehensive comparison of the fMRI encoding performance with different parameter choices, reported as the voxel-wise correlation on Subject 1. For experiments with different numbers of prototypes, we fix the sparsity constraint as $k=128$. For experiments with sparsity constraint, we fix the number of prototypes as $P=1000$.

\begin{figure}[htbp]
    \centering
    \begin{minipage}{0.49\textwidth}
        \centering
        \includegraphics[width=\linewidth]{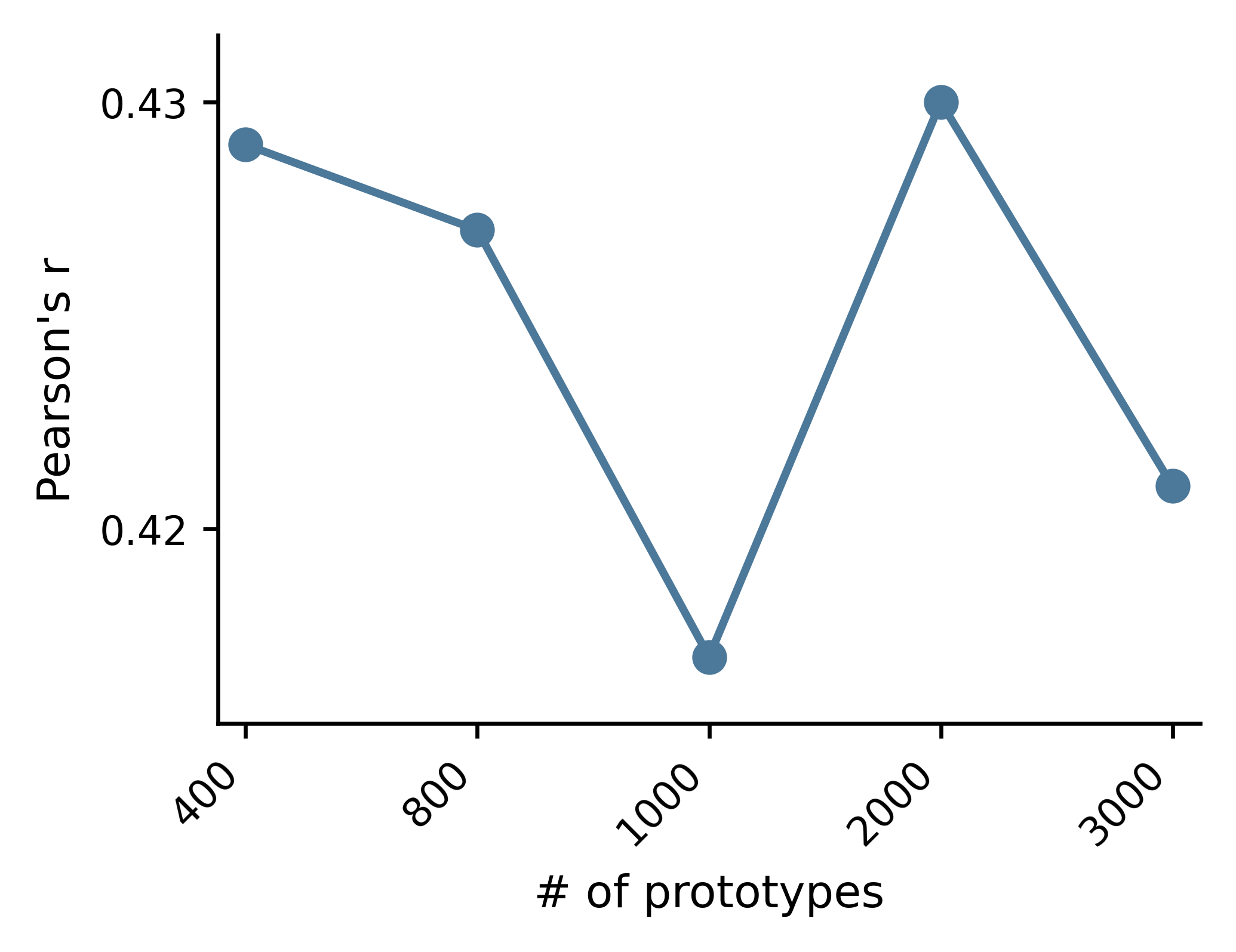}
    \end{minipage}
    \hfill
    \begin{minipage}{0.49\textwidth}
        \centering
        \includegraphics[width=\linewidth]{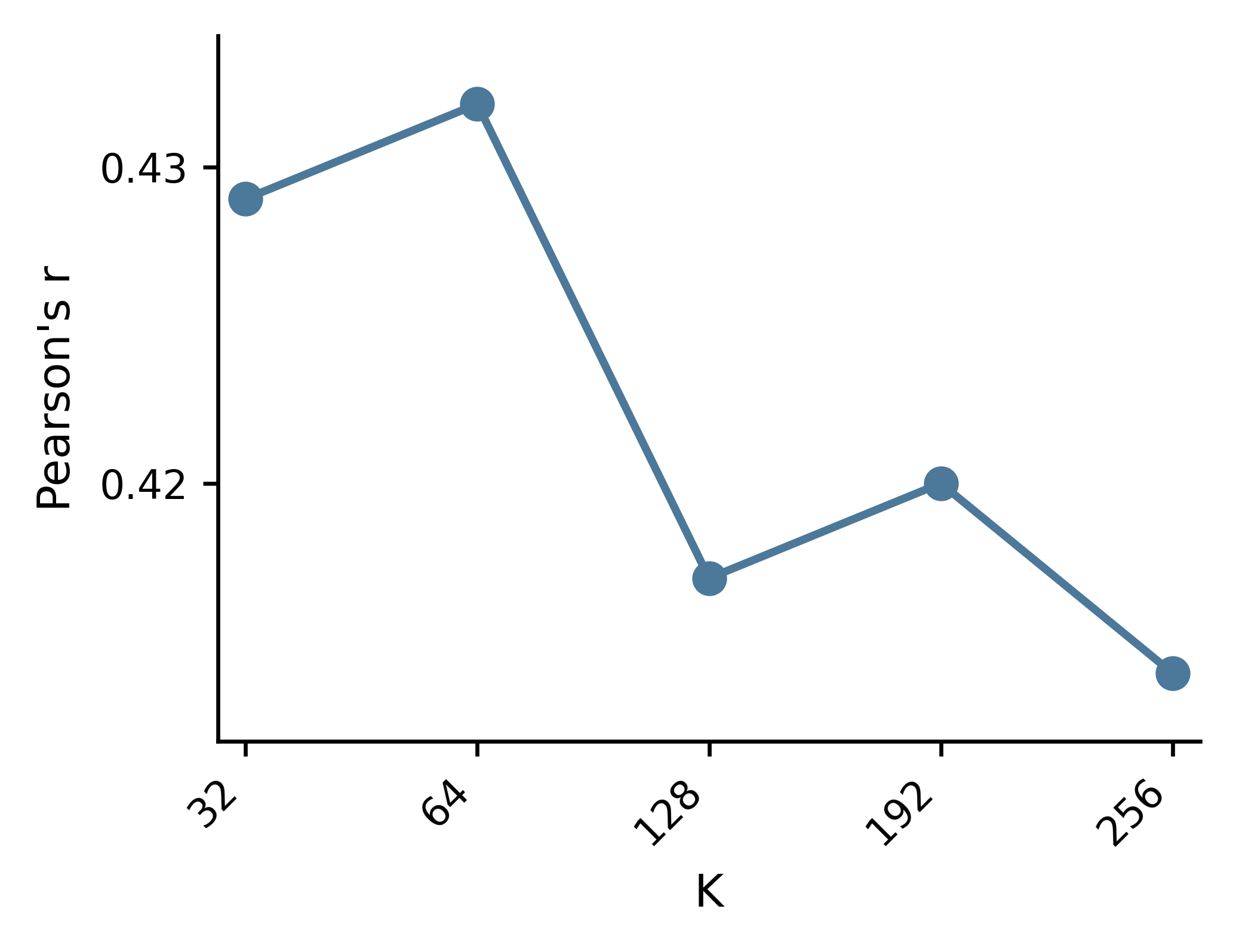}
    \end{minipage}
    \caption{fMRI encoding performance averaged across 12 different brain ROIs. For experiments with different numbers of prototypes (left), we fix the sparsity constraint as $k=128$. For experiments with sparsity constraint (right), we fix the number of prototypes as $P=1000$}
    \label{supfig:hyperparameter}
\end{figure}

Results in Figure \ref{supfig:hyperparameter} suggest that our approach is relatively robust to the selection of these two hyperparameters: increasing the number of prototypes or loosening the sparsity constraints leads to a minor drop of performance (\textit{i.e.,} around 0.01 drop in Pearson's correlation). Based on these observation, we choose sparsity constraint $k=128$ and $P=1000$ for a good balance between interpretability (\textit{i.e.,} reserving sufficient number of prototypes with reasonable granularity) and performance.

\section{Diffusion Model for Causal Intervention Experiment}
\label{sup:diffusion}
Different from studies \cite{luo2023brain,Takagi_2023_CVPR, feng2026towards, NEURIPS2023_mindeye, brainit} that rely on generative models pretrained with language data (\textit{e.g.,} Stable Diffusion \cite{Rombach_2022_CVPR} and CLIP \cite{clip}), we opt to train our own generative models from scratch with only visual data. This enables us to analyze the patterns of the visual cortex without assuming prior structures, \textit{e.g.,} without imposing language structure on representations in the visual cortex. 

\begin{figure}
    \centering
    \includegraphics[width=0.9\linewidth]{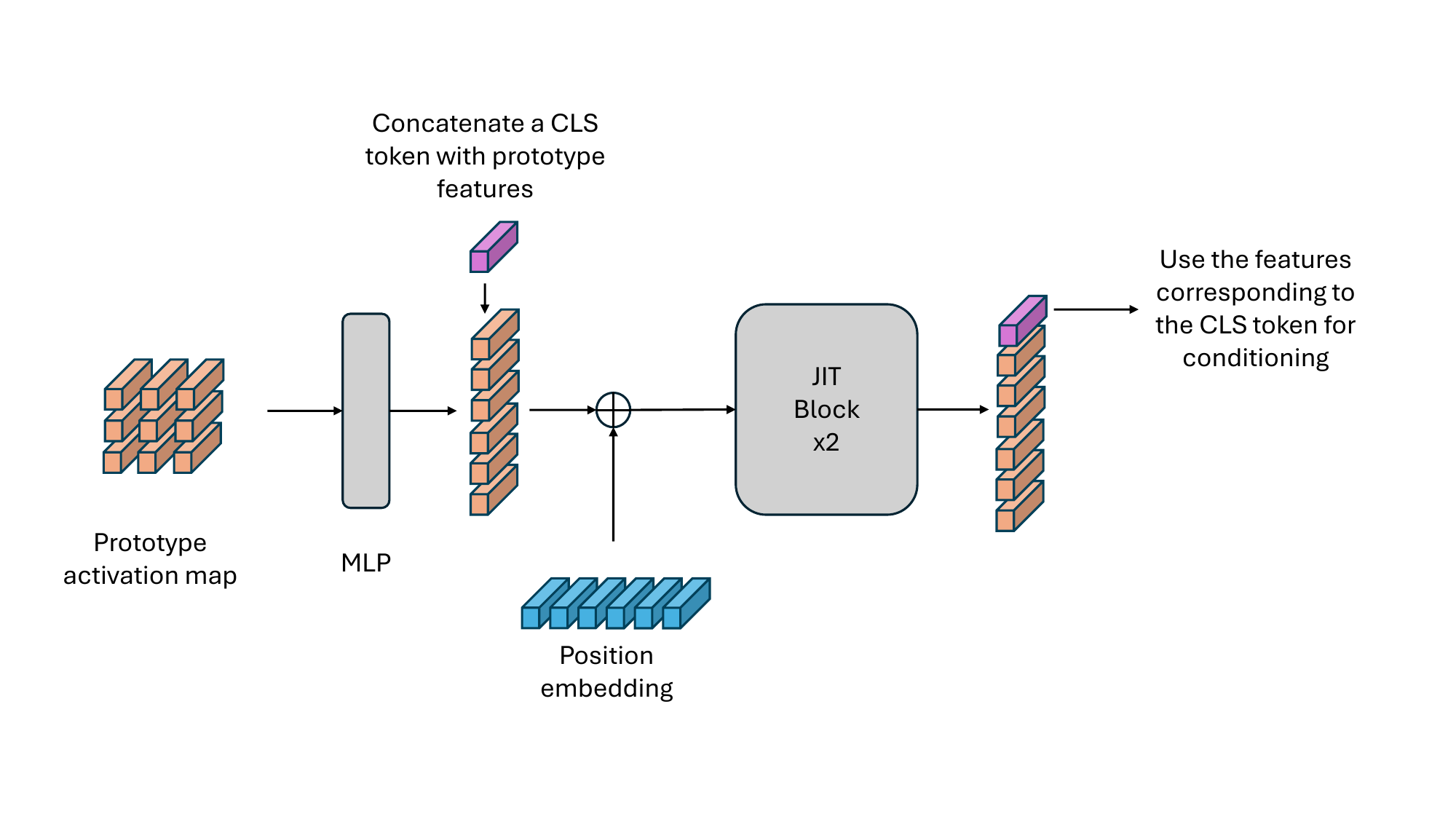}
    \caption{Schematic for the design of prototype encoder used in the diffusion models.}
    \label{supfig:diffusion}
\end{figure}

Specifically, as shown in Figure \ref{supfig:diffusion}, we choose the recently introduced JiT \cite{li2026basicsletdenoisinggenerative} diffusion model as our testbed, which follows a minimalist design while achieving state-of-the-art performance. While the original goal of JiT is to \emph{synthesize} images given class labels, here our objective is to \emph{reconstruct} images based on their prototype activation maps. Therefore, we substitute the class encoder with a prototype encoder (see the figure below), and also disable the classifier-free guidance \cite{ho2021classifierfree}. We train the model on ImageNet \cite{deng2009imagenet} with the standard parameters used for the original model, except that we reduce the noise level by half to facilitate learning of reconstruction.

\section{Experiment Details}
\label{sup:model_details}
\paragraph{Vision foundation model.} We use ViT-B16 and ConvNeXt-Base from DINOv3 \cite{dinov3}, and ViT-H from JEPA \cite{ijepa} as our backbone vision foundation models. For feature decomposition, we take the normalized patch features from the last layer.

\paragraph{Brain ROIs.} For this study, we focus on 12 established brain ROIs with distinct selectivity, including Occipital Face Area (OFA) and Fusiform Face Area (FFA-1 and FFA-2) selective for faces; Extrastriate Body Area (EBA) and Fusiform Body Area (FBA-1 and FBA-2) selective for bodies; Occipital Place Area (OPA), Parahippocampal Place Area (PPA), and Retrosplenial Cortex (RSC) selective for places; Occipital Word Form Area (OWFA) and Visual Word Form Area (VWFA-1 and VWFA-2) selective for words.

\paragraph{Visualization of prototypes.} To visualize the representative images for each prototype, we first identify the top-50 images with the highest activation averaged across all patches. To ensure consistency among the final representatives, we then compute pairwise cosine similarities between CLIP~\cite{clip} embeddings of these candidates and apply spectral clustering~\cite{ng2001spectral} to select 20 images that are both highly activated and representative of a shared visual concept associated with the prototype.

\paragraph{Optimization for prototypes.} We use Adam \cite{kingma2015adam} as our optimizer for training the prototypes (stage-1). The model is trained for 15 epochs with the official training and validation split from MSCOCO \cite{lin2014microsoft}. The learning rate is set to $10^{-3}$, with exponential decay. Batch size is set to 128. The best model is selected based on the validation performance.  

\paragraph{Optimization for fMRI encoding model.} Upon learning the prototypes, we extracted the prototype activation maps for images used in the NSD dataset \cite{nsd_dataset}. The same Adam optimizer is used to train the fMRI encoding model, which takes the fixed prototype activations and predicts the voxel responses. We create our own training, validation, and test set for each subject by splitting the official training set into $80\%/10\%/10\%$. We train our encoding model on a subject basis, and select the voxel from a single or multiple brain ROIs, depending on the analysis (\textit{e.g.,} for categorical selectivity in \S \ref{res:brain-roi-profile}, we train an encoding model for each brain ROI, while for prototype mapping in \S \ref{res:localization}, we train the model to predict responses for multiple ROIs.) The learning rate is set to $10^{-2}$, with batch size $32$, and number of epochs $20$.

\paragraph{Computational resource.} Our main model is relatively lightweight, and trained on a customized Linux server with two RTX 4090 GPUs (24GB VRAM per GPU). For the diffusion model used in our causal intervention analysis, we use another Linux server with one H100 and two RTX 6000 (96GB VRAM per GPU).

% \clearpage
% \newpage
% \input{checklist.tex}

\end{document}